\documentclass[useAMS,usenatbib]{mnras}
\usepackage{amsmath,amssymb,graphicx,color,longtable,enumitem,url,hyperref,multirow,rotating}

\title[Simulating time-varying strong lenses]{Simulating time-varying strong lenses}
\author[Vernardos]{G. Vernardos\thanks{E-mail: georgios.vernardos@epfl.ch} \\
Institute of Physics, Laboratory of Astrophysics, Ecole Polytechnique Fédérale de Lausanne (EPFL), Observatoire de Sauverny, 1290 Versoix, Switzerland
}
\date{Accepted XXX. Received YYY; in original form ZZZ}
\pubyear{2020}

\begin{document}
\label{firstpage}
\pagerange{\pageref{firstpage}--\pageref{lastpage}}
\maketitle

\newcommand{\h}{$H_{\mathrm{0}}$}
\newcommand{\rxj}{RXJ~1131-1231}

\begin{abstract}
We present a self-consistent and versatile forward modelling software package that can produce time series and pixel-level simulations of time-varying strongly lensed systems.
The time dimension, which needs to take into account different physical mechanisms for variability such as microlensing, has been missing from existing approaches and it is of direct relevance to time delay, and consequently \h, measurements and caustic crossing event predictions.
Such experiments are becoming more streamlined, especially with the advent of time domain surveys, and understanding their systematic and statistical uncertainties in a model-aware and physics-driven way can help improve their accuracy and precision.
Here we demonstrate the software's capabilities by exploring the effect of measuring time delays from lensed quasars and supernovae in many wavelengths and under different microlensing and intrinsic variability assumptions.
In this initial application, we find that the cadence of the observations and combining information from different wavelengths plays an important role in the correct recovery of the time delays.
The Mock Lenses in Time (MOLET) software package is available at: \url{https://github.com/gvernard/molet}
\end{abstract}

\begin{keywords}
	gravitational lensing: micro -- accretion, accretion discs
\end{keywords}

\section{Introduction}
\label{sec:intro}
Gravitational lensing refers to the deflection of light rays from a background source due to the presence of a massive object close to the line of sight.
In the case of strong galaxy-scale lensing, a galaxy-source (the source) will appear multiply imaged, magnified, and deformed into arc-like shapes, or even complete rings, around a galaxy-lens (the lens).
The lensed images generally remain static in time, unless the source happens to contain some time varying and bright astrophysical object, namely, either an Active Galactic Nucleus (AGN) that manifests as a quasar in the optical wavelengths or an exploding supernova.
In this case, gravitational lensing as a function of time has implications for measuring the Hubble constant, \h, a major and very timely cosmological application.

The current paradigm of the standard cosmological model, postulating a flat Universe consisting mainly of dark matter and dark energy \citep{Komatsu2011}, has been challenged by precise measurements of \h: different probes - the cosmic microwave background (CMB) radiation in the early Universe \citep{PlanckCollaboration2018vi} and the local cosmic distance ladder \citep[e.g.][]{Riess2016} - give values that disagree at the 3.4$\sigma$ level, which could imply the existence of new physics \citep{Freedman2017}.
Strong lensing provides an entirely independent physical probe to measure \h~under different assumptions and systematics \citep{Treu2016,Oguri2019}, potentially resolving the current crisis in cosmology \citep{deGrijs2017}.
Currently, there is an increased interest in understanding the systematic uncertainties in measuring \h~through lensed quasars \citep{Millon2020a} and supernovae \citep{Suyu2020}.
However, these efforts require robust measurements of time delays and good knowledge of the lens mass distribution: from the large scales of the order of several kpc, down to the distribution of stellar mass objects within the lens.
The former can be affected by several factors, like an incomplete description of the main lens mass-model \citep[e.g.][]{Birrer2019,Millon2020b}, the presence of substructure \citep[][]{Keeton2009,Gilman2020b}, and the mass density along the line of sight \citep[i.e. the mass-sheet degeneracy,][]{Rusu2017,Wong2017,Shajib2019}.
In the latter case, the presence of compact stellar-mass objects near the line of sight induces microlensing of point-like source light components \citep{Schmidt2010}, like the supernovae and quasar accretion discs of interest here\footnote{We disregard the possible, yet unlikely, scenario of a very bright star in the source crossing a caustic or a microcaustic.}, which can contaminate the observed light curves by introducing an additional extrinsic time variability \citep{Liao2019} and time delay \citep{Tie2018a}.

Apart from measuring \h, lensed quasars have two additional notable astrophysical applications: performing tomography of supermassive black hole immediate environments and detecting the presence and measuring the mass of dark massive substructures ($\sim10^7$ M$_{\odot}$ or less) around the lens \citep{Dalal2002,Metcalf2002} or along the line of sight.
In both cases, the time varying signal of the quasar and, most importantly, superimposed microlensing modulations play a crucial role.
In the first case, it is the onset of those rare microlensing high magnification events, most notably caustic crossings \citep[$\approx$1 per decade per system,][]{Mosquera2011b}, that carry the desired, but still elusive, information on the geometry and kinematics of quasar central regions under the effect of the extreme gravity of the supermassive black hole \citep{Padovani2017}.
Such measurements would complement reverberation mapping \citep[e.g.][]{Fausnaugh2017} and direct imaging by the Event Horizon Telescope\footnote{Which, however, will be able to directly image only 2-3 more nearby quasars, like it did for M87.} \citep{EHT2019} by a bigger sample of objects that are also at cosmological distances (having implications on quasar formation, black hole growth, galaxy evolution in general, etc).
In the second case, microlensing can mimic the effect of (de)magnification \citep{Schechter2002} caused by dark substructures on individual quasar images \citep[e.g.]{MacLeod2013,Nierenberg2017,Stacey2018} and produce similarly `anomalous' flux ratios that can be used to constrain dark matter \citep{Gilman2020a,Harvey2020,Chan2020}.
However, because there is a chromatic dependence of microlensing through the effective source size, its effect can be mitigated and/or disentangled by multiwavelength flux ratios.
Both of these applications are very timely because of the upcoming monitoring campaign of the Rubin Observatory Legacy Survey of Space and Time \citep[LSST;][]{LSST2009}, which is expected to discover up to a hundred microlensing events per year \citep{Neira2020}, and the Euclid space-based survey, which will provide high resolution multiwavelength snapshots, rich in microlensing signal \citep[][]{Vernardos2019a}.

Here, we present the Mock Lenses in Time (MOLET) software package that incorporates self consistently time variability of lensed quasars and supernovae, and we explore the effect of such variability on measuring time delays.
In order to report high precision measurements of \h, it is crucial to apply self-consistently our current understanding of the inter-dependency and degeneracies of all the physical processes involved in measuring the time delays.
So far, microlensing effects have been approximated to various degrees, or even ignored, due to their prohibitively large parameter space and associated cost of the required ray-shooting simulations \citep[with a few exceptions, e.g.][]{Bonvin2018,Chen2018,FoxleyMarrable2018,Goldstein2018b}.
In this work, we take advantage of the GERLUMPH microlensing simulations and software tools \citep{Vernardos2014a, Vernardos2014b} to create a fully self-consistent pixel-level forward-simulation tool, which integrates all the related aspects of lensing and observations in a coherent way.
In Section \ref{sec:physical_processes}, we outline the various physical systems and model components of relevance and in Section \ref{sec:implementation} the implemented forward-simulation framework.
Section \ref{sec:results} showcases a novel application of MOLET: measuring time delays and their uncertainties from lensed quasars and supernovae with different observations.
We present our conclusions in Section \ref{sec:discuss}.

\section{Description of physical processes}
\label{sec:physical_processes}
Here we present the different physical processes that become relevant at different space and time scales.
Emphasis is given on their inter-dependencies and the limitations and degeneracies of related models - we do not attempt to provide a complete review on lens theory and applications.
Clarifications and specific references are provided whenever necessary, while the reader is referred to the reviews by \citet{Keeton2010} and \citet{Schmidt2010}, the lecture notes of \citet{Schneider2006}, or any other textbook on gravitational lensing for more details.

\subsection{Time-independent lensing}
\label{sec:macromodel}
The lens mass model, or macromodel, refers to the mass distribution of the main deflector - a two-dimensional projected mass distribution on the plane of the sky (under the thin lens approximation) - and it can explicitly include nearby companion galaxies and substructures (in the same lens plane), and/or more subtle tidal effects as external shear.
The corresponding lens potential determines a number of key physical properties, namely the deflection angles (the first order derivatives of the potential), the magnification matrix (the Jacobian of the potential), and the light arrival-time surface, which are all a function of the position on the plane of the sky.
The latter has a number of extreme points (viz. a maximum, minima, and saddle-points) that define the locations where multiple images of a point-source would appear (according to Fermat's principle) and their difference in arrival-time, i.e. the time delays.
The deflection angles create the overall appearance of the lensing system by acting on the extended source brightness profile and creating lensed arcs, rings, and multiple extended images - in the following we refer to these features collectively as `extended lensed features' and use the term `multiple images' to refer exclusively to the point-source.
Finally, the magnification matrix describes the change in shape and total area of a surface element from the source to the lens plane, thus determining the magnification of each multiple image.

The magnification matrix can be expressed as a function of the total shear vector, $\boldsymbol{\gamma}$, and the total convergence, $\kappa$, fields, with the latter being in fact the surface mass density.
Convergence is split between two components: one that is smooth across all scales, corresponding to dark matter and gas within the lens, and a grainy component in the form of compact stars and stellar remnants.
Compact matter on the macroscopic (kpc) scale can be described by a smooth field as well, however, at scales comparable to the point-source size (spanning angles of $\approx10^{-6}$ arcsec on the sky) it consists of distinct objects that act as microlenses (point-mass lenses).
Hence, these three fields, $\boldsymbol{\gamma}$, $\kappa$, and the smooth-to-total matter fraction, $s$ (or the compact matter convergence, $\kappa_{\star}$), fully describe, in a statistical sense, any subsequent (time-dependent) microlensing variability of the multiple images\footnote{The orientation of the shear vector determines the angle of elongated microlensing caustics, increasingly present as the magnitude of the shear grows. This direction is important with respect to the point-source velocity because the source may happen to move along a trough, or valley, between high magnification regions of clustered elongated caustics. Such troughs are devoid of any microlensing caustics and therefore do not induce any high magnification, rather demagnify the source.}.

Mass distributions along the line of sight to the lens and source and at different redshifts can further act as lenses/sources.
Such a scenario, however, requires the close alignment of three or more massive astrophysical objects along a given direction through the Universe, which, although less frequent than the already rare usual single deflector lensing, has been observationally confirmed \citep[see][for the recent discovery of an impressive third source in the J0946$+$1006 lensing system]{Collett2020}.
Here we focus on the time-variability of the lenses, leaving substructure effects along the line of sight \citep{Despali2018} and multiplane lensing \citep{Gavazzi2008,Petkova2014} as future extensions to our method.

\subsection{Time variability}
\label{sec:time_variability}
The intrinsic physical mechanisms responsible for the time variability of quasars and the light profile evolution of supernovae are quite complex and sometimes poorly understood.
Their time-variable observational signatures, however, are well studied and can be described by empirical models.
In the following, we always consider a given intrinsic variability light curve of the source at its rest-frame in the absence of lensing.

The collective effect of tens to thousands of microlenses on a point-source is described by a magnification map.
Such maps describe the total increase/decrease in flux from the sum of the areas of the created micro-images (unresolved multiple images of the point-source, separated by $\approx10^{-6}$ arcsec, whose surface brightness is conserved) as a function of source plane position, and consist of intertwined networks of caustics: lines separating high from low magnification regions, corresponding to a higher or lower number of micro-images \citep[always by an increment of two, see][]{Schneider2006}.
If any part of the source, which can now have finite dimensions when compared to the caustics, overlaps with any such line, there is an increase in magnification; the higher the fraction of total source flux contained by the part touching the caustic, the more dramatic the magnification \citep[with the extreme being a true point-source, which has an infinite magnification on the caustic, see][]{Fluke1999}.

The number of microlenses per unit area in the lens plane is directly related to the local value of $\kappa_{\star}$, and is either set to the sum of equal mass microlenses or the integral (weighted sum) of a mass function\footnote{For all practical purposes, microlensing is always taking place within length scales far too small for $\kappa_{\star}$, $\kappa$, and $\boldsymbol{\gamma}$, to diverge from their values at the location of each multiple image.}.
The microlens positions are practically unobservable, hence each map results from a random realization of a microlens ensemble.
As long as the map is wide enough, it can be considered as an accurate statistical representation - \citet{Vernardos2013} find a value of 25 Einstein radii\footnote{The Einstein radius is the typical lensing scale length, or caustic size, of a single lens, depending only on the lens mass and the angular diameter distances to the source and the lens. For an ensemble of lenses, the mean mass is used.} to be adequate.
In practice, magnification maps are produced as high resolution pixellated versions of caustic networks, using variations of the numerical inverse ray-shooting technique \citep{Kayser1986}.

The position of the source on the magnification map is another unobservable, random variable.
Nonetheless, the source is not static: the relative velocities of the observer, source, and lens\footnote{Although there is a velocity vector associated with the proper motion of each microlens, which causes the caustic pattern to move leading to ``faster'' variability \citep[e.g.][]{Kundic1993}, its magnitude becomes significant only in very low redshift systems.
\citet{Poindexter2010a} explicitly include such a component in their calculations, while \citet{Wyithe2000a} have shown that this effect can be approximated by a bulk velocity for all the microlenses based on the lensing galaxy's velocity dispersion.}, result in a motion of the source on a trajectory across the magnification map defined by an effective velocity vector on the source plane \citep[][]{Kayser1986,Kochanek2004,Neira2020}.
In addition to the source displacement, there can be a variation in the source effective size for reverberating quasars and expanding supernovae.
Hence, different parts of the source can undergo a varying degree of magnification, which can be up to a few magnitudes over a period of months to years, and produce a time-dependent signal as they cross or expand into neighbouring regions, filled or devoid of caustics.
This signal is superimposed on the intrinsic source signal and independent between the multiple images (extrinsic).
The different possible such sources that can be found in lensing systems are:

\begin{enumerate}[label=(\roman*),leftmargin=*]
	\item \textit{Moving disc.} This is the most commonly used quasar microlensing variability model \citep[e.g.][]{Morgan2012,Morgan2018,Cornachione2020}. It consists of a fixed accretion disc brightness profile that is crossing the magnification map at a given velocity. The intrinsic variability is assumed to take place simultaneously across the entire disc, scaling its brightness up or down.
	\item \textit{Moving and reverberating disc.} The quasar instrinsic variability and its light profile are, in fact, connected through physical mechanisms and geometry. For example, when a temperature variation is produced at the center of the accretion disc, it propagates outwards with a finite speed, changing the brightness of the regions it crosses with a time lag with respect to the center. This is the so-called `lamp-post' model \citep{Cackett2007}, which, in combination with microlensing, has been found to introduce an additional time delay \citep{Tie2018a}. The brightness profile will be time dependent but not with a simple overall scaling, as in the previous case, rather with a time-lagged brightening of different regions of the accretion disc, which are also differently microlensed. The net effect is a time varying brightness profile that is crossing the magnification map.	
	\item \textit{Moving disc with un-microlensed component.} This model ignores the response of the innermost regions of the accretion disc to central variations but includes a delayed emission coming from the much larger Broad Line emission Region \citep[BLR, e.g.][]{Sluse2014}. The latter one is much less affected by microlensing but the delayed intrinsic variability shapes the eventual signal by introducing more small-scale structure in the light curves (Paic et al. 2021, in prep.).
	\item \textit{Expanding supernova.} This is the most commonly used supernova microlensing variability model \citep[e.g.][]{FoxleyMarrable2018,Goldstein2018b}. The main assumption is that the supernova expansion velocity ($\sim10^4$ km/s) is much higher than the effective source plane velocity ($\sim10^2$ km/s), therefore, the source center-of-light can be considered static. In this case, microlensing variability arises due to different regions of the supernova light profile overlapping with different magnification regions of the map as it expands. The expansion velocity and light profile geometry can vary as a function of time.
	\item \textit{Expanding and moving supernova.} In some cases, the above assumption might not be true: in systems having a maximum observer velocity \citep[CMB dipole transverse velocity, see figure 1 of][]{Neira2020}, a low redshift ($z_{\mathrm{L}} \lesssim 0.15$) and a high-mass lens, the mean effective source plane velocity can be of the order of 10 per cent of the supernova expansion velocity. This can be considerably higher if the lens and source peculiar velocities happen to align parallel to the observer's velocity as well. Although less probable, such a system is plausible and requires a simultaneous expansion and translation of the source across a magnification map to be taken into account.
\end{enumerate} 

In all the above scenaria, the most crucial structural parameter of the microlensed point-source is its size, expressed by the half-light radius, $r_{\mathrm{1/2}}$.
It has been shown that the detailed geometry of the source brightness profile does not play any significant role in bulk microlensing properties \citep[][although the opposite holds for short-lived high magnification events]{Mortonson2005,Vernardos2019b}.
If the source is smaller than a single magnification map pixel then values from the map can be used directly (subject to shot noise from the ray-shooting simulations).
Such a situation is not optimal for expanding or slow-moving sources though, because if there is a caustic present, its fine structure is washed-out over the entire pixel and hence a higher resolution map is required to correctly capture the induced variability.
In the case of the source being larger than a map pixel, then a convolution of the map with the light profile is required before extracting any magnification values.

\section{Simulation technique and implementation}
\label{sec:implementation}
Each macromodel determines the values of $\kappa,\gamma,s$ at the locations of the multiple images, which, in turn, define the statistical properties of a magnification map.
Computing a single map requires ray-tracing simulations with deflections due to hundreds to tens of thousands of microlenses - a computationally demanding task \citep{Bate2012}.
Although there are several forward-simulation techniques to generate static lenses \citep[see the review by][]{Plazas2020}, none of these include the simultaneous self-consistent production of microlensing observables, viz. flux ratios and light curves, precisely due to the prohibitive associated computational cost.

The novelty of the approach presented here is the self-consistent inclusion of microlensing through pre-computed magnification maps and software acceleration.
The GERLUMPH\footnote{\url{http://gerlumph.swin.edu.au}} parameter survey provides a systematic coverage of the entire $\kappa,\gamma,s$ parameter space of interest with high-quality readily available magnification maps \citep{Vernardos2014a, Vernardos2014b}.
These maps have a sufficiently large width (25 Einstein radii), high resolution (400 pixels per Einstein radius), accurate statistics ($>10^9$ light rays shot per map), and are scalable with the mean microlens mass\footnote{GERLUMPH maps have been created using a fixed mass for the microlenses. Although distributing the baryonic mass to an ensemble of stellar mass objects using a given Initial Mass Function prescription is physically justified, it has been shown to have a minimal effect \citep{Wambsganss1990b,Lewis1995,Wyithe2001}.}.
Software acceleration by Graphics Processing Units, which enabled GERLUMPH, is also critical in performing convolutions between maps and finite-sized point-sources, leading to speed-ups of $\times10$ \citep{Vernardos2015}.
This is particularly relevant in the case of expanding sources, which require a convolution of a differently sized profile at each time step (as opposed to a fixed source moving across a map that requires only one convolution).
The software library \textsc{GERLUMPHPP}\footnote{\url{https://github.com/gvernard/gerlumphpp}}, provides an easy interface to harness GPU acceleration (but still compatible with CPUs), and has already been used in a number of microlensing applications \citep{FoxleyMarrable2018,Neira2020}.

Our simulation technique does not depend on any specific assumption on lens or source populations and instrument specifications.
Theoretical or empirical relations between different physical models/properties can be provided independently, e.g. one could relate the mass with the light profile of a fiducial elliptical lens galaxy using the fundamental plane, make further assumptions on the mass-to-light ratio or the total and dark matter mass profile to determine $\kappa_{\star}$, assume different point-source light profile, geometry, effective velocity, intrinsic variability, etc.
A basic choice of specific physical models is provided, e.g. mass, light profile, etc, however, our method can easily be extended by additional user-provided functions and data.
A flowchart describing the simulations' pipeline is shown in Fig. \ref{fig:flowchart}; below we summarize its various components:

\begin{figure*}
	\includegraphics[width=\textwidth]{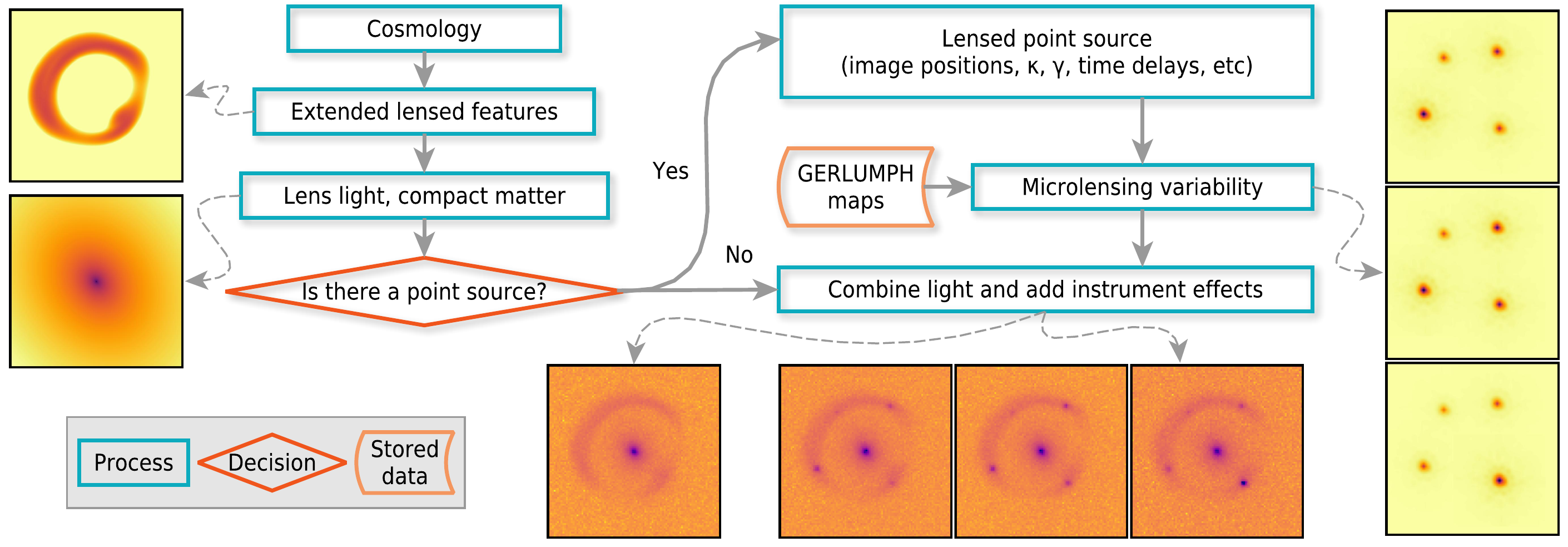}
	\caption{Flowchart of the MOLET software package.}
	\label{fig:flowchart}
\end{figure*}

\begin{enumerate}[label=(\roman*),leftmargin=*]
	\item \textit{Cosmology.} We calculate the angular diameter distances to the lens, source, and between them, based on a given cosmological model. The value of \h~needs to be provided here.
	\item \textit{Lens.} The following properties of the deflector are required:
		\begin{itemize}[leftmargin=*]
			\item Mass model: it can consist of a single, or several analytical models \citep[see][for a catalogue]{Keeton2001b}, or pre-computed lens potential values, $\psi(x,y)$, which, for example, can be based on numerical N-body hydrodynamical simulations \citep[e.g.][]{Mukherjee2018}. An external shear vector can be included as well. We currently provide the Singular Isothermal Ellipsoid \citep[SIE;][]{Kormann1994} and softened power-law \citep{Barkana1998} models.
			\item Light profile: it can have one or more analytical components [here we provide S\'{e}rsic \citet{Sersic1963} and Gaussian profiles], or be an actual image of a galaxy. In the latter case, it should have a resolution higher than the simulated instrument to avoid introducing artifacts during the ray shooting and binning to the final pixel size.
			\item Compact mass model: in the case of a point-source, such a model is required to calculate $\kappa_{\star}$ at the location of its multiple images. It can be an analytical profile (we provide S\'{e}rsic and Gaussian), or a mass-to-light ratio value to convert the light profile directly into compact stellar mass.
		\end{itemize}
	\item \textit{Extended source.} Any light profile, similarly to the lens light profile, can be used for the source. Its location with respect to the macromodel caustics determines the final appearance of the lensing system, i.e. the number of multiple extended images and if they are merged into arcs or a ring.
	\item \textit{Point source.} The location of the point-source with respect to the macromodel caustics determines the number of multiple images and their configuration, i.e. cross, cusp, or fold. The half-light radius, $r_{\mathrm{1/2}}$, is the main parameter of the point-source's actual light profile, while its detailed shape becomes relevant only with respect to high magnification events, observed with a very fine temporal resolution ($\sim$day). A mean mass value of the compact objects acting as microlenses is required to set the Einstein radius and consequently the microlensing magnification map length scale. Here we use GERLUMPH maps, but any magnification map could be used, e.g. generated according to a microlens mass function, as long as it is converted to the compatible format first. Regarding the evolution of the point-source with respect to the microlensing magnification map caustics (translational movement or expansion), we can have the following cases:
		\begin{itemize}[leftmargin=*]
			\item \textit{Moving disc}. We use the effective source plane velocity model from \citet{Neira2020}, which consists of: a fixed velocity vector for the observer based on the CMB velocity dipole that is a function of the system's sky coordinates, a randomly directed vector of the combined peculiar velocities of the lens and the source with magnitude drawn from a corresponding Gaussian distribution \citep[eq. 5 of][]{Neira2020}, and a random velocity for the microlenses equal in magnitude to the velocity dispersion in the lens.
			\item \textit{Moving and reverberating disc}. We use the `lamp-post' quasar variability model \citep{Cackett2007} to relate a central temperature variation to a brightness profile variation that propagates outwards through the accretion disc. For an analytic light profile, one needs to compute its first derivative and relate it to the central variability via a time lag \citep[see][]{Tie2018a}. Hence, at each time step a different light profile is produced, which is then microlensed as it moves across a magnification map. Here we provide a lamp-post variability model for the thin disc \citep{Shakura1973}.
			\item \textit{Moving disc with un-microlensed component.} We use an additive non-microlensed intrinsic flux for each multiple image \citep[see eq. \ref{eq:signal} here and 4 of][]{Sluse2014}.
			\item \textit{Expanding supernova}. We use a fixed expansion velocity and shape for the supernova light profile, but our method can be easily extended to include time varying velocities and profile shapes (also for the moving model below).
			\item \textit{Expanding and moving supernova}. We use the same effective velocity model as a moving disc and combine it with a supernova expanding light profile.
		\end{itemize}
	\item \textit{Instrument.} The following characteristics of existing or hypothetical instruments need to be specified:
		\begin{itemize}[leftmargin=*]
			\item Resolution: the pixel size of the detector's camera.
			\item Field of view: the dimensions of the simulated patch of sky.
			\item PSF: the point-spread function.
			\item Observing strategy: a sequence of dates when the simulated system was observed. This can be related to an assumed RA and DEC of the simulated system, combined with a specific facility's observing schedule (e.g. see call for white papers on LSST cadence optimization\footnote{\url{ls.st/doc-28382}}).
			\item Noise: here we provide a simple uniform Gaussian noise model calculated from a given signal-to-noise ratio. Existing simulators of varying complexity can also be used \citep[e.g][for LSST's camera]{Lage2019}.
		\end{itemize}
\end{enumerate}

To produce the extended lensed images, a grid of pixels is assumed on the lens plane and then deflected backwards on to the source plane using the deflection angles computed from the lens mass model.
Once the location of the deflected rays on the source plane is determined, the lensed brightness values are computed by evaluating an analytical source profile or interpolating between the associated adjacent pixels of a numerical one.
At the same time, the critical and caustic lines of the mass model are computed, which are used in validating the multiple image configuration.

To find the multiple images of a point source at a given location, we cover the lens plane completely with a coarse grid of triangles, which we then deflect on to the source plane using the lens mass model.
Finding which triangles contain the point source gives an approximate location of its multiple images.
We then redefine a finer grid of triangles around these approximate locations and repeat the process until the size of the grid becomes smaller than 1/100 of the final simulated instrument's resolution.
The final output is the positions of the multiple source images and an associated uncertainty, which is below 1 per cent.

The intrinsically variable flux of the point-source needs to be provided as an input \citep[through simulations, e.g. see][for quasars and supernovae respectively]{Sartori2019,Kessler2009}, and has to span at least $t_{\mathrm{obs}} + \Delta t_{\mathrm{max}}$, where $t_{\mathrm{obs}}$ is the duration of the observing period and $\Delta t_{\mathrm{max}}$ the maximum time delay.
External microlensing variability is calculated separately and consists of a varying multiplicative magnification factor for the time period $t_{\mathrm{obs}}$.
The final point-source signal for each multiple image q is then:
\begin{equation}
\label{eq:signal}
F_{\mathrm{q}}(t) = M_{\mathrm{q}} \mu_{\mathrm{q}}(t) F_{\mathrm{in}}(t-\Delta t_{\mathrm{q}}) + M_{\mathrm{q}} F_{\mathrm{rev}}(t-\Delta t_{\mathrm{q}}),
\end{equation}
where $M$ is the constant macro-magnification, $\mu(t)$ the microlensing magnification, $\Delta t$ the time delay, and $F_{\mathrm{in}}$ the intrinsic source flux.
The additive term is only valid in the presence of an un-microlensed flux component (more pertinent to quasar sources), $F_{\mathrm{rev}}$, that can be modelled by a response of the BLR to the central variability:
\begin{equation}
\label{eq:rev}
F_{\mathrm{rev}}(t) = f_{\mathrm{BLR}} \left[ \Psi(t,\tau) \circledast F_{\mathrm{in}}(t) \right],
\end{equation}
where $\Psi$ is a transfer function, $\tau$ a time lag usually corresponding to the light travel time from the central regions to the BLR, and $f_{\mathrm{BLR}}$ the fraction of this flux that is eventually re-emitted in the same wavelength.

The three simulated light components, viz. extended lensed features, point-source, and lens light, are created first in a super-resolved grid, having $\times10$ smaller pixels than the final simulated instrument's resolution.
Convolution with the PSF takes place in this grid, over-sampling the PSF to match the smaller pixel size if necessary.
For the point-source light, we simply scale an image of the PSF (convolution with a delta function) according to the time-varying signal from eq. (\ref{eq:signal}) for each multiple image.
The images are then binned to the instrument's resolution and noise is added\footnote{We note that the convolution and binning operations are not necessarily permutable and reversing their order may overly smooth out the lensed images or, conversely, introduce spurious pixel edge effects.}.

The most critical observational parameter with respect to time-variable lenses is organizing their monitoring, i.e. the frequency and distribution of the observations over time.
\citet{Neira2020} have shown the effect of different observing (LSST-specific) strategies in relation with detecting high magnification events in lensed quasars and \citet{Goldstein2018b} with respect to lensed supernovae from ZTF and LSST.
While factors like instrumental effects, noise, and artifacts, do affect the resulting simulated images as every other astronomical observation, modelling them correctly, which depends on various physical mechanisms affecting incoming photons from the lensed system reaching Earth's atmoshpere (or Earth's orbit) and finally being recorded by an instrument's detector, is of secondary importance for the purposes examined here.
In any case, our method can be easily extended by additional instrumental modules, consisting of a model for the PSF, noise, and artifacts.

The essential final products of the above procedure are the continuously (daily) sampled light curves of each multiple image.
Based on this, one can study the effect of different observing strategies for each science application.
In addition, our method produces pixel-level simulations of cutout images of lens systems at different time intervals.
Such images are necessary to calibrate methods that analyze the actual observational pixel data, from deblending the lensed images to lens mass modelling.

\subsection{The MOLET software package}
The independent simulation components, viz. extended lensed features, point-source image locations, microlensing, etc, are reflected in the modular design of the MOLET software package outlined in Fig. \ref{fig:flowchart}.
Each procedural component is a C++ program written using the \textsc{GERLUMPHPP} and \textsc{VKLLIB}\footnote{\url{https://github.com/gvernard/vkl_lib}} libraries for microlensing and strong lensing respectively.
Minimal shell scripts are used to manage some operating system specific tasks (e.g. checking and creating files), images are written in \textit{.fits} format, and data in \textit{.json} files, easily shared between the different modules.
The code is open source and each module can be replaced by any other program that performs the same scientific computations, as long as the output data format is adhered to.
All models presented above are implemented in this first version of the software, except a moving and reverberating disc and an expanding and moving supernova (both cases of a moving and variable in size source).
These are planned to be included in a future version, together with a self-consistent way to get a reverberated flux component from the BLR (eq. \ref{eq:rev}) and possibly a basic intrinsic variability model \citep[like the damped random walk, e.g.][]{MacLeod2010}.
All third-party software required to run MOLET is provided in software containers, easily deployable in any operating system or supercomputer.
The source code and detailed instructions can be found at the project's homepage\footnote{\url{https://github.com/gvernard/molet}}.

\begin{figure}
	\includegraphics[width=0.5\textwidth]{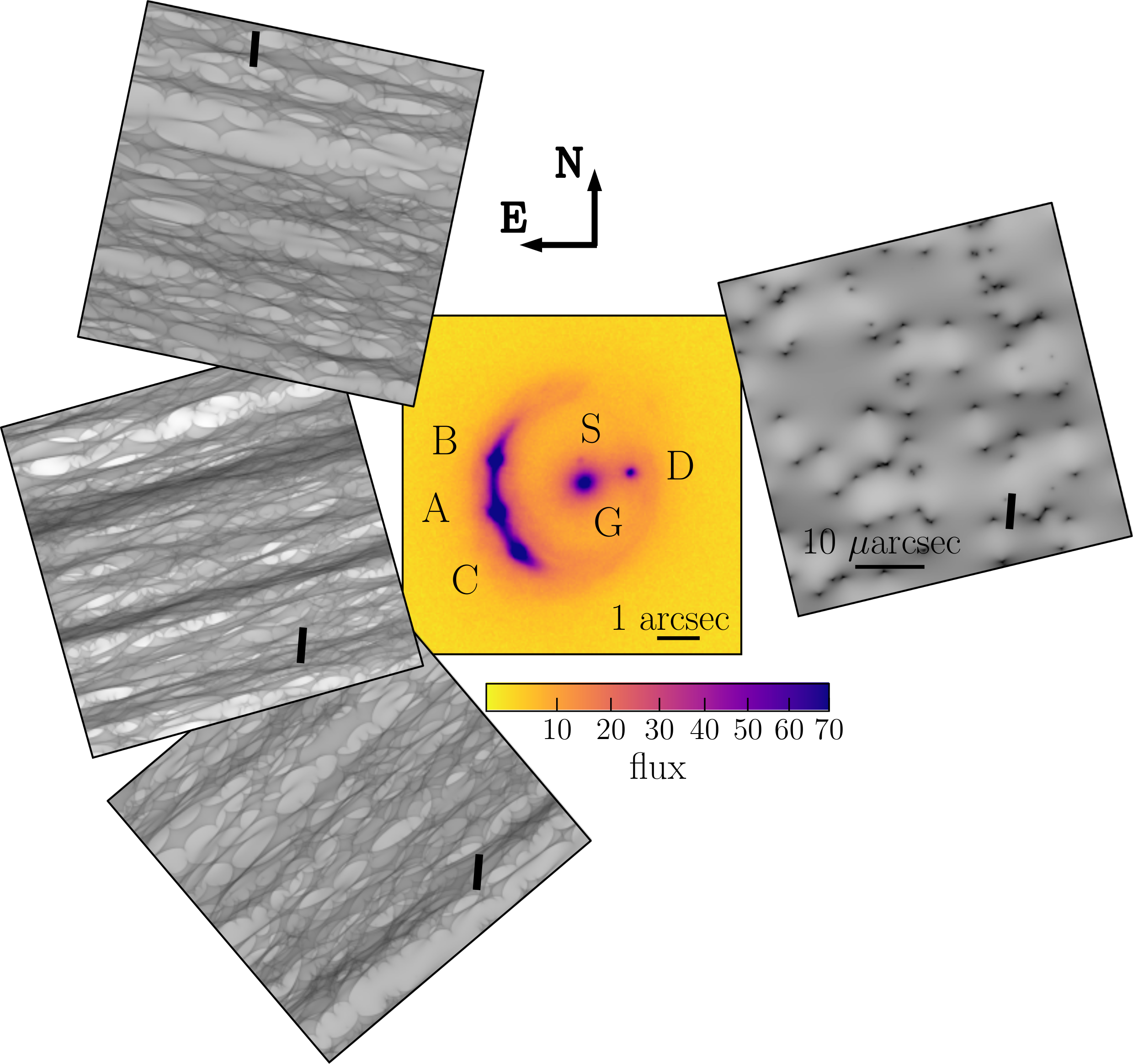}
	\caption{Mock image of \rxj~and associated magnification maps at the locations of the quasar multiple images. The maps are consisted with the mass model, i.e. have the corresponding $\kappa,\gamma,s$ (see Table \ref{tab:macro}), as well as being aligned with the local total shear vector. A trial trajectory for the quasar point source is shown with an effective velocity drawn from the distribution given in fig. 3 of \citet{Neira2020}. Its length has been multiplied by a factor of 5 to better visualize that the trajectories in each map are parallel - constituting, in fact, just a single trajectory on the source plane where the different map realizations are projected. The corresponding microlensing light curves are shown in Fig. \ref{fig:theo_light_curves}.}
	\label{fig:configuration}
\end{figure}

\begin{figure}
	\includegraphics[width=0.5\textwidth]{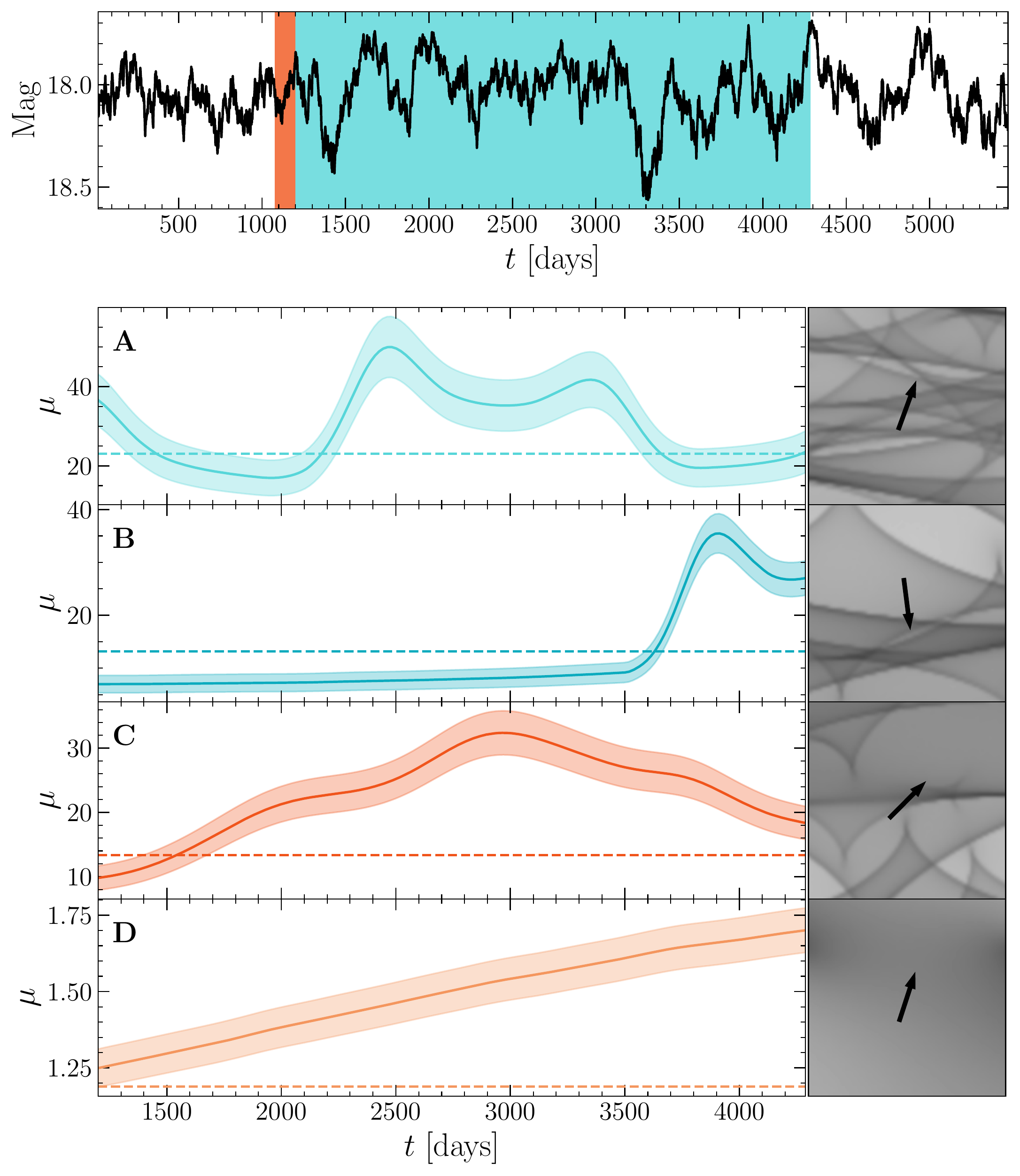}
	\caption{Top: assumed intrinsic variability of the quasar point source of \rxj. The blue-shaded area shows the $\approx$10-year time-window of the assumed monitoring campaign and the orange-shaded one the time window into the past provided by the maximum time delay of the system (corresponding to the unobservable maximum image that has a delay of $\approx150$ days). Bottom: microlensing magnification for each image, and corresponding zoomed-in region of the magnification maps with the source trajectory indicated by the arrow (same as the trajectories in Fig. \ref{fig:configuration}). The horizontal dashed lines correspond to the macro-magnification of each image, listed in Table \ref{tab:macro}. The shaded areas indicate the Poisson error of the magnification in the GERLUMPH map pixels, resulting from the numerical inverse ray-shooting technique \citep{Kayser1986} that was used to generate them.}
	\label{fig:theo_light_curves}
\end{figure}

The novelty of MOLET is employing microlensing simulations in real time and self-consistently.
This is enabled by the $>75,000$ pre-computed GERLUMPH magnification maps that cover extensively the $\kappa,\gamma,s$ parameter space of relevance.
Running MOLET locally requires to have copies of the required magnification maps; when such copies are not present a link to download them is provided automatically.

Generating microlensing light curves is the most computationally intensive task.
In particular, if several convolutions between maps and source profiles are required, as is the case for expanding or time-varying source light profiles, the computational cost on the CPU can become prohibitive.
However, GPUs can accelerate this task by a factor of 20 \citep{Vernardos2014b} and MOLET can use the GPU-compiled version of \textsc{GERLUMPHPP} to achieve this.

\section{Results}
\label{sec:results}
To demonstrate the capabilities of MOLET, we present simulated monitoring of the quasar \rxj \citep{Sluse2006} over a period of $\approx$10 years.
This particular object is interesting because it is known to undergo microlensing, it is being currently monitored by the COSMOGRAIL program \citep{Millon2020b}, and it is extensively used to derive \h~from time delay studies \citep[e.g.][]{Suyu2013,Wong2020}.
Here, we use the results of \citet{Chen2016} for the parameters of the lensing potential, the lens light profile, the quasar point source location, the reconstructed light profile of the extended source, and the reconstructed PSF (see their tab. 1, fig. 9, and fig. 10 respectively).
Consequently, the simulated data are for the Keck Adaptive Optics K$^{\prime}$ band, which has a pixel-scale of 0.04 arcsec \citep[see fig. 1 of][]{Chen2016}.
A mock image for the system is shown in Fig. \ref{fig:configuration}.

\begin{figure*}
	\includegraphics[width=\textwidth]{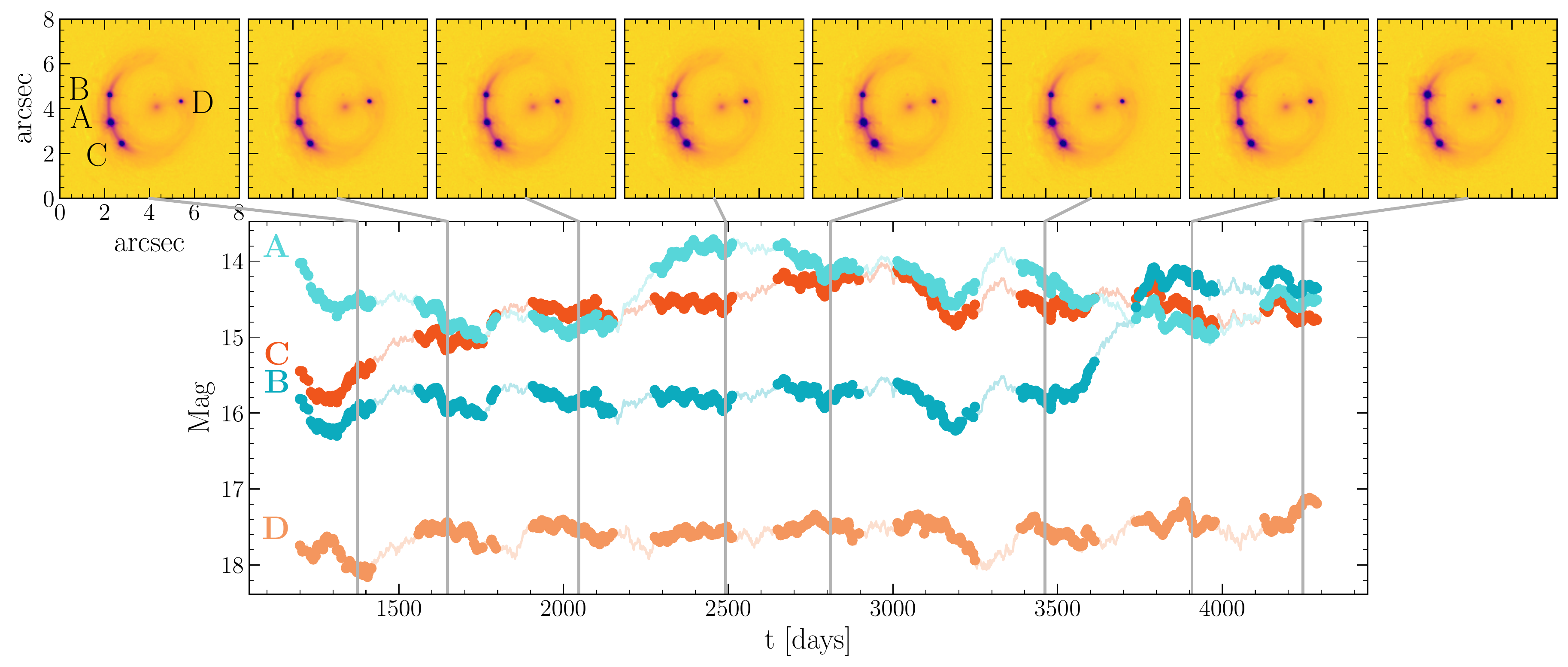}
	\caption{Example of continuous (lines) and sampled (points) mock light curves and corresponding mock Keck K-band cutouts of \rxj. The sampling of the light curves is the same as the one from the COSMOGRAIL observations. The cutouts are shown for illustration purposes only and have not been used to extract the light curves. Here we have scaled the lens and source brightness and skewed the color coding of the pixels to better show the variability of the multiple images, e.g. a clear brightening of image A due to microlensing is observed around 2500 days.}
	\label{fig:film_strip}
\end{figure*}

For the quasar point-source, we use the coordinates of its recovered position on the source plane\footnote{G.C.F. Chen, private communication.} to find the corresponding multiple images created by the lens potential.
MOLET currently requires the intrinsic variability of the quasar to be provided as input, which we get as a realization of a Damped Random Walk \citep[DRW, see][for a justification of this choice]{MacLeod2010} with the characteristic timescale and variance\footnote{We use the values of 80 days and 13 for these two parameters - D. Sluse, private communication.} parameters derived from the observed light curve of \rxj \citep[][]{Sluse2014}.
A realization of an intrinsic light curve is shown in the top panel of Fig. \ref{fig:theo_light_curves}.
Finally, there is a secondary point source visible in the Keck AO data, attributed to a star forming region of the quasar host galaxy, which, however, we omit, as monitoring of such faint point sources is unrealistic with current instruments.
Nevertheless, dealing self-consistently with (the rare case of) multiple point sources and their associated microlensing can be an easily implemented extension of MOLET.

The local values of $\kappa$ and $\gamma$ for the multiple quasar images result directly from the lens potential.
For $s$, one needs to provide a distribution for the projected compact (stellar) mass of the lens.
Usually, the mass is assumed to follow the lensing galaxy's light distribution with some mass-to-light ratio.
Proceeding in this way would require the calibration of the fluxes reported in \citet{Chen2016} to standard filters.
Instead, and because our goal is not to study the actual distribution of stars within the lens, we assume a compact mass profile identical to the main S\'{e}rsic light component of the lens but scale its amplitude using a reasonable value for the mass density in stars, in this case $\rho_{\rm eff} = 1.0$ kg/m$^2$.
The magnification maps at the locations of the multiple images, shown in Fig. \ref{fig:configuration}, are directly retrieved from GERLUMPH, with their $\kappa,\gamma$, and $s$ values matching the ones resulting from the lens macromodel as closely as possible (see Table \ref{tab:macro}).
These square maps are 25 (10000) Einstein radii (pixels) wide.
We set a uniform mass of 1 M$_{\odot}$ for the microlenses.

\begin{table}
	\centering
	\caption{Properties of the multiple images located at the coordinates $x$ and $y$ on the image plane (in arcsec). The shear angle, $\phi_{\gamma}$, is given in degrees east-of-north and the time delay, $\Delta t$, is given in days with respect to image B. $\mu$ is the macro-magnification. The last three rows show the closest $\kappa,\gamma,s$ values from GERLUMPH.}
	\label{tab:macro}
	\begin{tabular}{rcccc}
		&A&B&C&D\\
		\hline
		&\multicolumn{4}{c}{Macromodel}\\
		x&2.27&2.23&2.76&5.38\\ 
		y&3.38&4.6&2.45&4.31\\ 
		$\kappa$&0.49&0.44&0.46&0.93\\ 
		$\gamma$&0.55&0.49&0.46&0.92\\ 
		$\phi_{\gamma}$&-164.44&-11.75&-139.72&-166.47\\ 
		$s$&0.72&0.69&0.71&0.76\\ 
		$\mu$&-23.07&13.19&13.36&-1.19\\ 
		$\Delta t$&0.8&0&0.2&94.6\\
		\hline
		&\multicolumn{4}{c}{GERLUMPH}\\
		$\kappa$&0.49&0.44&0.46&0.95\\ 
		$\gamma$&0.55&0.49&0.46&0.91\\ 
		$s$&0.7&0.7&0.7&0.8
	\end{tabular}
\end{table}

For the accretion disc and its effective velocity across the source plane, we adopt the same models as \citet{Neira2020}, whose parameter values are given in their table 1.
The half-light radius of the accretion disc as a function of wavelength is given by a simple power-law model with a slope equal to $4/3$, i.e. the same as the thin-disc model; for an observing wavelength of 2190 nm (the midpoint of the standard photometric K filter) the resulting half-light radius of the disc is $7.47 \times 10^{14}$cm, or 0.016 times the value of the microlenses' Einstein radius on the source plane.
The light profile of the disc is assumed to have the shape of a normal distribution viewed face-on.
The fiducial effective velocity that produces the trajectories on the magnification maps shown in Fig. \ref{fig:configuration} has a magnitude of 294 km/s, which is very close to the expected mean, and a $-126$ deg direction east-of-north \citep[see fig. 3 of][]{Neira2020}.
Except image D, for which it is rare to lie close to a caustic (see the corresponding magnification map in Fig. \ref{fig:configuration}), all the other images have light curves that cross dense caustic regions on the maps, producing the rich in features light curves shown in Fig. \ref{fig:theo_light_curves}.

For the time sequence of the observations we choose 600 observing dates from the COSMOGRAIL light curve of \rxj, which have intervals of roughly once a week and season gaps, with an arbitrary starting point (only the relative time intervals are relevant).
Photometric errors can be added directly to the light curves according to instrument specifications, depth, etc, as we will perform below, or calculated from the simulated cutouts through deblending and light curve extraction techniques.
An example of resulting simulated light curves taking into account all the variability effects and the macroscopic time delays is shown in Fig. \ref{fig:film_strip}.

\begin{figure}
	\includegraphics[width=0.5\textwidth]{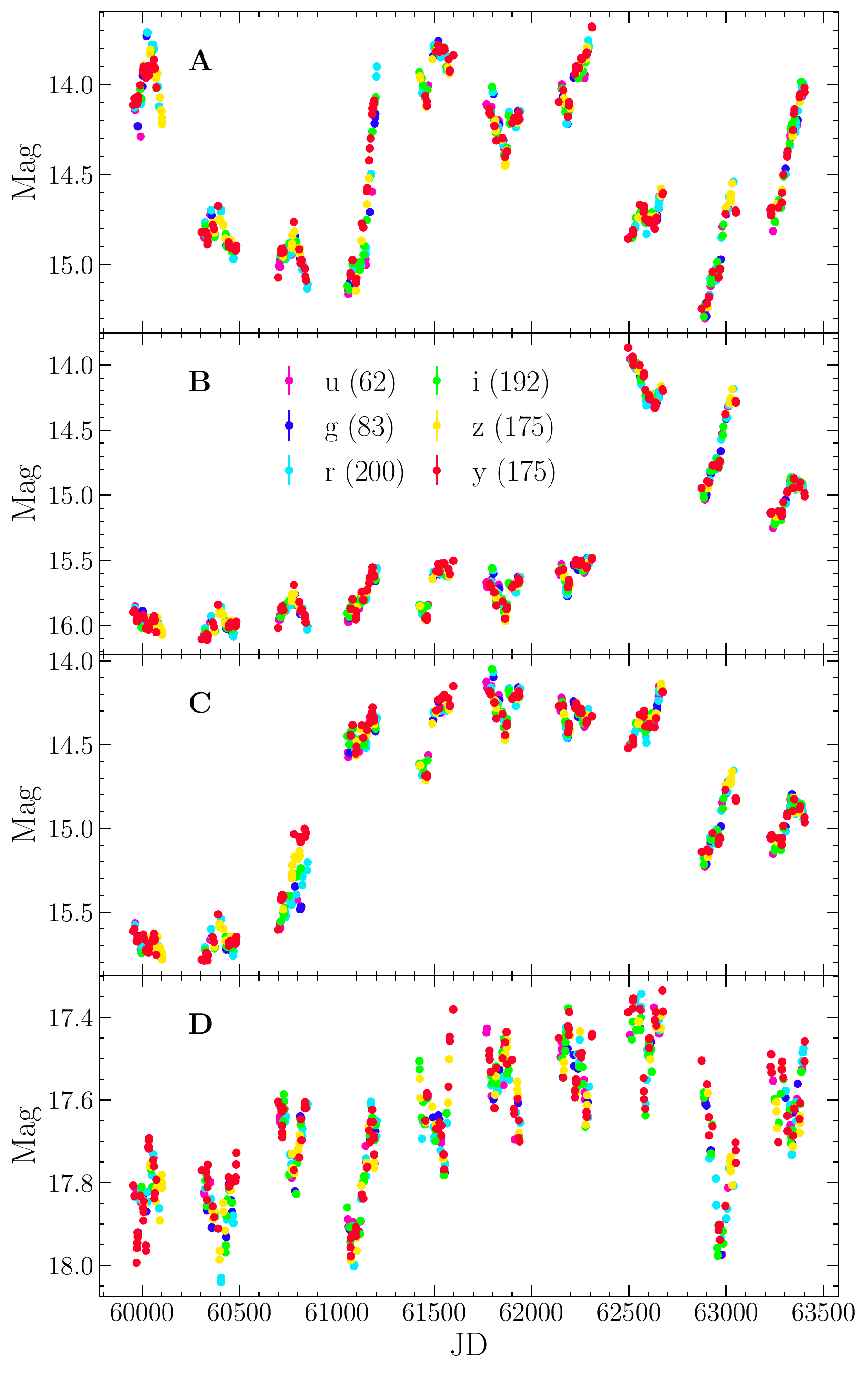}
	\caption{Example of LSST-observed light curves corresponding to those shown in Fig. \ref{fig:film_strip}. Here, the smaller accretion disc gives rise to sharper high magnification events. The colossus\_2664 LSST observing strategy was used in this particular example to determine the time intervals between observations in each filter \citep[see][for more examples]{Neira2020}. The numbers in the legend indicate the number of data points in each band. The error bars of 0.005 mag are smaller than the size of the data points.}
	\label{fig:lsst_lcs}
\end{figure}

We can also simulate multiwavelength light curves for \rxj, like the ones expected in the 6 LSST filters.
After choosing a specific observing strategy \citep[see][who examined several observational strategies for this system]{Neira2020}, the only input to determine the dates of each visit at the patch of sky containing the lens are its coordinates.
The accretion disc model is the same as the one used above, leading to similarly small sizes: in the u band the source is smaller than a magnification map pixel and in the y band the source is still $\approx 3$ times smaller than the one assumed in Figs. \ref{fig:theo_light_curves} and \ref{fig:film_strip}.
For simplicity, we keep the same intrinsic light curve across the 6 filters, although smoother (sharper) intrinsic variability is expected for longer (shorter) wavelengths.
The differences in disc size for our model are of the order of a light day, too small to produce significant variations in the intrinsic light curves anyway.
An example of resulting multi-wavelength light curves is shown in Fig. \ref{fig:lsst_lcs}, where we do observe the expected microlensing behaviour: the bluer (redder) the wavelength the sharper (smoother) the features.

\begin{figure*}
	\includegraphics[width=\textwidth]{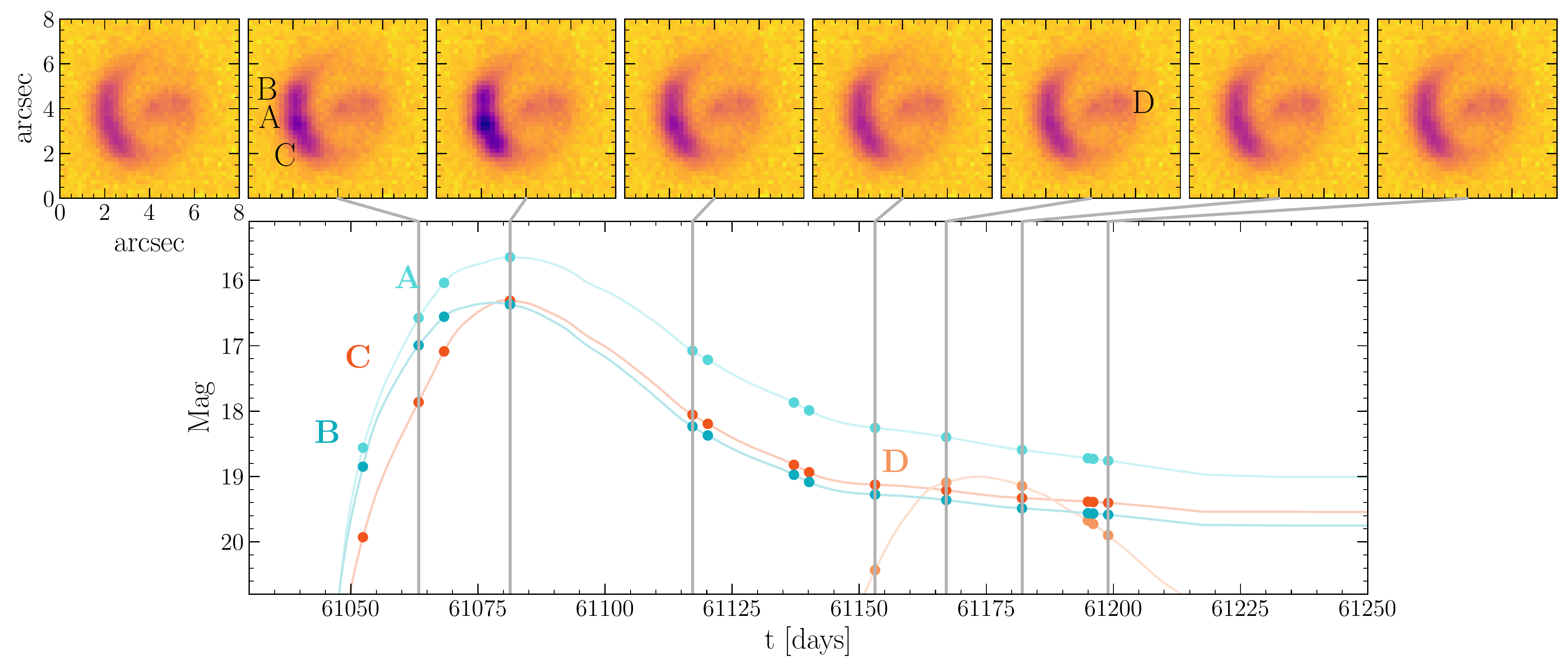}
	\caption{Example of continuous (lines) and sampled (points) mock light curves and selected cutouts in the LSST i band, for a hypothetical system having exactly the same mass and light configuration as \rxj, but having replaced the quasar source with an exploding supernova. The observing dates result from the the colossus\_2664 LSST observing strategy. Here we have scaled the brightness of the supernova host galaxy and skewed the color coding of the pixels to better show the variability of the multiple images. Image D becomes noticeable around JD61165. The first cutout shows the lens before the supernova explosion (without any point-source light).}
	\label{fig:film_strip_SN}
\end{figure*}

Finally, we can also use MOLET to produce mock lensed supernovae light curves.
As an example, we keep the same lens configuration but change the nature of the source from a variable quasar to an exploding supernova.
We simulate the observations with the same LSST observing strategy as in the previous case.
We assume a brighter than typical Type Ia supernova that peaks at 16 mag in the i band, having the geometry of a uniform face-on disc that expands with a constant velocity of $10^5$ km/s and with the same size in each wavelength.
This corresponds to a size range from $<10^{15}$ cm ($<0.02$ Einstein radii) on day one, about 8 map pixels, to $1.6\times10^{17}$ cm (4 Einstein radii) after 186 days, when we assume that the phenomenon stops\footnote{Long term microlensing effects in supernovae have been very little explored so far.}.
We generate the intrinsic supernova LSST light curves from the software package \textsc{SNCosmo} \footnote{\url{https://sncosmo.readthedocs.io/en/stable/}}, and assume an arbitrary date for the onset of the explosion.
Fig. \ref{fig:film_strip_SN} shows an example of such simulated light curves in the LSST i band, for a single microlensing realization.

\subsection{Measuring time delays}
\label{sec:time_delays_quasars}
Measuring time delays in each of the above observational setups has its own challenges and requirements; here, we assess the effect of different intrinsic and microlensing variability realizations.
We use resolved light curves, not cutouts, therefore the only relevant instrumental parameters are: for quasars, the wavelength of the observations that sets the corresponding accretion disc size, and for supernovae, the cadence of the observations.
We keep the same lens potential, point source position, and time delay distance \citep[1935 Mpc,][]{Chen2016}, therefore the same time delays, listed in Table \ref{tab:macro}.
For a flat Lambda Cold Dark Matter cosmological model with $\Omega_\mathrm{m}=0.3$, the time delay distance corresponds to a value of 84 km/s/Mpc for \h.
Our focus here is to assess the uncertainty and bias of techniques that measure the time delays, not \h, hence such a high value is not a problem \citep[although not too surprising either for this specific lens, e.g.][]{Wong2020}.
The measured COSMOGRAIL time delay between the leading image B and image D - the most meaningful to examine since the other delays are too small, between 1 and 2 days - is $93.7^{+2}_{-2}$ days \citep{Millon2020a}, practically the same as the $94.6$ days we get under the assumptions of our model.

First, we consider \rxj~observed in the r band of the ECAM \citep[$600-725$nm, see][for details]{Millon2020b}.
This results in an accretion disc $\approx6$ times smaller than the one in the Keck K band used in the examples shown in Figs. \ref{fig:theo_light_curves} and \ref{fig:film_strip}.
Because of this smaller size, we expect more pronounced short-range microlensing features to be present.
The cadence of the observations is the same as in the example shown in Fig. \ref{fig:film_strip} and photometric errors are added by hand; we use the relative errors from the publicly available COSMOGRAIL light curve for \rxj~and assign them to the simulated data points of each image in the same order in time but with an arbitrary starting point.
For the intrinsic variability, we assume the same DRW model and parameters as before and create 10 realizations with a mean apparent magnitude of 18, roughly matching the Euler telescope images in the r band.
For microlensing, we also create 10 realizations by sampling the effective velocity magnitude and direction.
We combine these using just the first term in eq. (\ref{eq:signal}) to create a set of 100 light curves to measure the time delays from.
A second set of another 100 simulations is produced, this time including an unmicrolensed intrinsic flux component from the BLR (the second term in eq. \ref{eq:signal}).
This flux is calculated from the same intrinsic light curve each time, assuming a top-hat transfer function, a time lag of 17 days from the size of the BLR \citep{Mosquera2011b}, and an eventual 20 per cent of the reverberated unmicrolensed flux making it into our r band filter \citep[see][in prep., for a similar application to the microlensed quasar Q~J0158-4325]{Paic2021}.

\begin{figure*}
	\includegraphics[width=\textwidth]{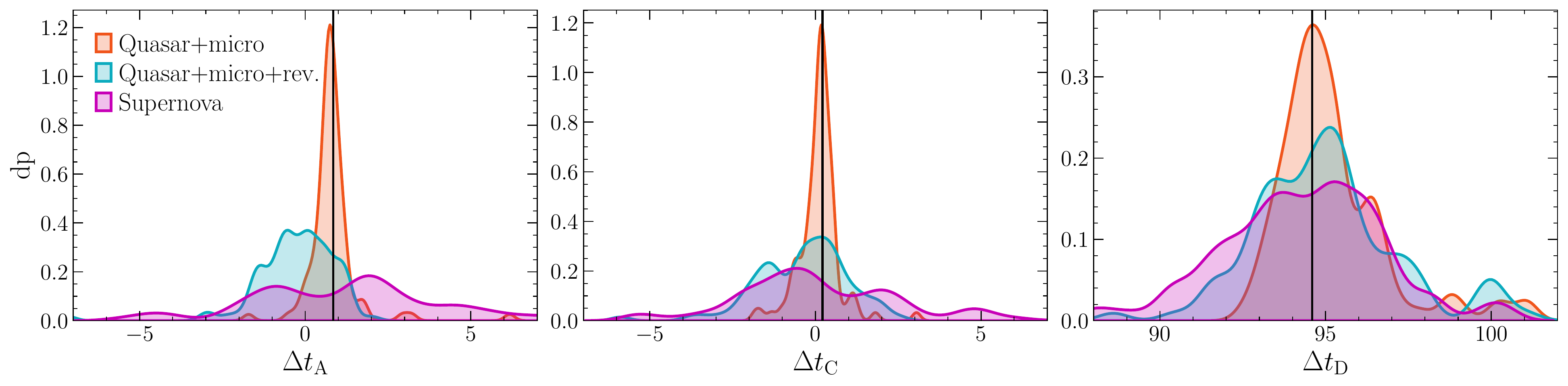}
	\caption{Probability density of measured time delays for different samples of simulated light curves. The two samples of 100 simulated quasar light curves each, one with intrinsic and microlensing variability only (orange) and one including an unmicrolensed reverberated flux component (turqoise), give a very similar distribution of measurements, albeit slightly broader for the case with reverberation. A sample of 100 simulated multiband supernova light curves, including microlensing, has a somewhat wider spread than the quasar measurements for image D (the other images having too small a delay of 1-2 days). The means and $1\sigma$ confidence intervals of the distributions are listed in Table \ref{tab:delays}, while the true values of the delays are listed in Table \ref{tab:macro} - indicated here by the vertical lines.}
	\label{fig:time_delays}
\end{figure*}

To measure the delays, we use the \textsc{PyCS}\footnote{\url{https://cosmograil.gitlab.io/PyCS3/}} software \citep{Tewes2013} to fit the light curves assuming spline polynomials to describe the intrinsic and microlensing signal and get a point estimate of the delays for each mock \citep[see][for a description of the different fitting methods available]{Millon2020c}.
In order to avoid convergence of the fit to some local minimum, we always initialize the fitting algorithm with the true delays rounded to the nearest day.
Finally, we assume a small number of knots in the splines to account for short timescale and high amplitude variability - as expected due to the small accretion disc and reverberated flux.
In Fig. \ref{fig:time_delays} we show the measured time delays for each image with respect to the leading image B.
The measurements are quite accurate, nicely clustering around the true values even for the short time delay images A and C, but not precise, having a spread of up to $\pm5$ days.
The mean measurements and 1$\sigma$ confidence intervals are listed in Table \ref{tab:delays}, where we can see that the presence of an unmicrolensed reverberated component decreases the precision for image D by roughly 1.5 days.

We replace the quasar with a supernova and create a set of 100 LSST light curves with different microlensing (we keep the same intrinsic variability).
To facilitate the convergence of the fitting algorithm, we have increased the LSST photometric errors to the fixed value of 0.005 mag.
We use the \textsc{SNTD}\footnote{\url{https://sntd.readthedocs.io/en/latest/}} software package to measure the time delays in 4 (r,i,z,y) of the 6 LSST bands, because u and g end up not having enough data points in the given time period.
A simultaneous fit across these bands is performed using the SALT-2 \citep{Guy2007} supernova variability model (here we ignore the fitted parameters of the intrinsic light curve and only keep the time delays).
The measured time delays for a sample with 100 different microlensing realizations are shown in Fig. \ref{fig:time_delays} and the mean and $1\sigma$ confidence intervals listed in Table \ref{tab:delays}.
For image D, the spread of these measurements is very similar to the one for a microlensed quasar with an unmicrolensed flux component.
However, it is important to note that LSST will not have an adequate cadence to constrain time delays with supernovae, hence the uncertainty in the measurements shown in Fig. \ref{fig:time_delays} is expected to reduce to 1-2 days when using dedicated follow-up \citep{Huber2019}.

\begin{figure*}
	\includegraphics[width=\textwidth]{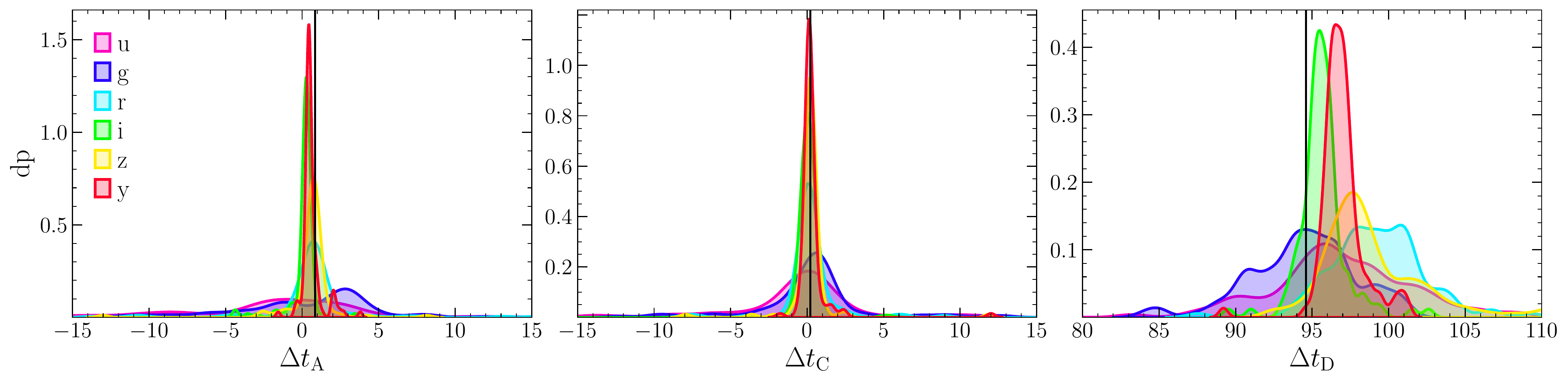}
	\caption{Probability density of time delays measured from a sample of 100 simulated light curves in each LSST filter. The number of data points in each light curve in each filter is shown in the legend of Fig. \ref{fig:lsst_lcs}. The means and $1\sigma$ confidence intervals of the distributions are listed in Table \ref{tab:delays}, while the true values of the delays are listed in Table \ref{tab:macro} - indicated here by the vertical lines.}
	\label{fig:lsst_time_delays}
\end{figure*}

\begin{table}
	\centering
	\caption{Measured mean time delays (in days) and $1\sigma$ confidence intervals from different datasets of 100 light curve realizations each.}
	\label{tab:delays}
	\begin{tabular}{rrrr}
		mock dataset &$\Delta t_{\mathrm{A}}$&$\Delta t_{\mathrm{C}}$&$\Delta t_{\mathrm{D}}$\\
		\hline 	
		Quasar+micro&$0.8_{-0.3}^{+0.4}$&$0.1_{-0.4}^{+0.3}$&$94.9_{-1.0}^{+1.5}$\\
		Quasar+micro+rev.&$-0.1_{-1.2}^{+0.9}$&$-0.2_{-1.4}^{+1.1}$&$94.8_{-1.7}^{+2.2}$\\
		Supernova&$1.4_{-2.7}^{+2.5}$&$-0.2_{-1.8}^{+2.6}$&$94.3_{-2.3}^{+2.1}$\\
		LSST u&$-1.1_{-7.0}^{+3.5}$&$-0.0_{-2.9}^{+1.7}$&$96.5_{-3.7}^{+4.7}$\\
		LSST g&$1.7_{-5.4}^{+1.8}$&$0.5_{-2.0}^{+0.9}$&$94.5_{-3.6}^{+2.9}$\\
		LSST r&$0.7_{-0.7}^{+0.6}$&$0.1_{-0.4}^{+0.6}$&$99.2_{-2.8}^{+2.2}$\\
		LSST i&$0.3_{-0.3}^{+0.2}$&$0.0_{-0.1}^{+0.1}$&$95.6_{-0.9}^{+0.8}$\\
		LSST z&$0.7_{-0.4}^{+0.3}$&$0.2_{-0.2}^{+0.2}$&$98.0_{-1.7}^{+4.4}$\\
		LSST y&$0.5_{-0.2}^{+0.4}$&$0.1_{-0.1}^{+0.3}$&$96.8_{-0.8}^{+1.0}$\\
	\end{tabular}
\end{table}

Finally, we repeat the experiment for the lensed quasar \rxj~using 100 LSST light curve realizations and measure the time delays separately in each filter with the same setup for \textsc{PyCS}.
We show our results in Fig. \ref{fig:lsst_time_delays}, where we observe that the accuracy and precision vary considerably per filter.
For images A and C, filters u and g have the largest spread, which could be explained by sparser observations and increased short-scale variability due to the smaller accretion disc, while the delays are well recovered in the other filters.
For the longest and most interesting delay of image D, the i and y bands have the smallest uncertainty, albeit with a slight, 1-2 day systematic offset towards longer delays.
This result is harder to explain because it does not correlate neither with the size of the accretion disc, nor with the number of observations.
The mean measurements and 1$\sigma$ intervals are shown in Table \ref{tab:delays}.

\section{Discussion and future prospects}
\label{sec:discuss}
We have presented a self-consistent and versatile forward modelling software package that can produce time series and pixel-level simulations of time-varying strongly lensed systems.
In summary, the main input required is a mass model for the lensing galaxy, including some assumption on its compact (stellar) matter distribution, and a variability model for the point-source (intrinsic and microlensing).
This, together with the observing wavelength and cadence, is sufficient to generate time series.
For simulated images, in addition to the above, the light profile of the lensing galaxy, the extended light from the source, and the observing instruments' resolution, PSF, and noise properties are also required.
The resulting images and time series can be used independently of each other - in the latter case, error bars will need to be provided separately by the user.
Finally, we note that images for static lenses can be generated as well by switching off the time-dependent part, however, the originality and wider applicability of the method stem from the latter.

There are no assumptions on specific populations or scaling relations inherent in the method, meaning that a very broad combination of various models and parameters is possible.
Although this allows for unrealistic combinations, and the responsibility of avoiding this is left to the user, it also allows for maximum flexibility.
For example, there is no constraint (or safeguard) on matching an extended source's brightness to its redshift, or the intrinsic mean brightness of a quasar to its accretion disc size.
Specifically, in this first version of the software all the intrinsic variability light curves and their relations (e.g. via transfer functions in case of reverberation, or spectral features of supernovae across wavelengths) need to be computed externally.
However, all the physical mechanisms that are connected within the context of strong lensing and microlensing are modelled self-consistently, e.g. the underlying cosmological model, the convergence and shear fields required for microlensing, conservation of extended source flux, etc.

Here we used a detailed mass model of \rxj~from the literature \citep{Chen2016} as a workbench to measure time delays with different methods and/or data.
Although we restricted the extent of the parameter space explored - this is a demonstrative study anyway - useful conclusions can still be drawn.
We find evidence that more regular monitoring of lensed quasars, like in the COSMOGRAIL program ($\approx$60 observations per year), performs better in measuring time delays than LSST (approx. between 6 to 20 observations per year depending on the filter), as shown in Table \ref{tab:delays}.
Cadence, however, is not the main limiting factor of the LSST measurements (see Fig. \ref{fig:lsst_time_delays}) and the role of other factors is yet to be explored.
Different components of the variable signal coming from the quasar (e.g. different relative power of high and low frequencies) seem to play a role in the precision of the measured time delays, as indicated by the broader measurements recovered here while assigning an additional 20 per cent of the intrinsic flux to an unmicrolensed component reverberated by the BLR.
Replacing the quasar with a supernova, leads to similar measurements, but in this case this is driven by the low number of data points ($\approx15$ per filter, see Fig. \ref{fig:film_strip_SN}), which should not be the case if an early alert for a supernova transient is provided.

To further understand the impact of different physical mechanisms and observing parameters on such time delay measuring experiments, we propose the following systematic explorations of the parameter space:
\begin{enumerate}[label=(\roman*),leftmargin=*]
	\item Investigate the effects of different intrinsic variability, cadence, and length of the observing period of lensed quasars.
	\item Athough it seems that the presence of a high frequency reverberated flux component does not bias the delays by much, different realizations of such variability will need to be explored to better understand its effect on time delays.
	\item The effect of uncertainties in the mass model parameters of the lensing galaxy is reflected in microlensing through the $\kappa,\gamma$. Similarly, $s$ is accompanied by assumptions on the stellar mass distribution and mass-to-light ratio. These uncertainties can be incorporated by sampling a distribution of these parameters when generating mock sets of light curves. Most importantly, this will allow to estimate the covariance between, for example, the mass model and the accuracy and precision of the measured time delays.
\end{enumerate}
For all these tests, independent information and physically motivated priors and covariances between parameters can and should be assumed or tested.
Finally, in the specific case of LSST, a time delay measuring algorithm fitting simultaneously data in many filters could increase the performance.

In our study, we have used only mock time series data, to which we arbitrarily introduced some observationally motivated error bars according to the assumed instrument, and we did not make any use of the mock cutout images other than in illustrative examples.
However, using directly the cutouts, e.g. within light curve extraction pipelines, could simultaneously allow for testing/improving these pipelines and for having real instrument error bars.
This could be achieved by, for example, injecting mock systems in telescope images at the early stages of data reduction and performing the same procedures on them as on the real lenses.
As a next step, high-resolution space-based images could be used to derive lens models, which in turn would be used to create several mock datasets in order to achieve model-aware source deblending and light curve extraction.
This is particularly useful in cases where multiple images are too close (e.g. merging pairs in fold configuration lenses, or low mass lenses), or when the extended source is bright enough to contribute a significant flux component at the locations of the multiple images.

We have demonstrated how our software can be combined with other methods, in this case time-delay measurement codes, in a streamlined fashion.
Specifically, we envisage our open-source modelling software to be used in connection with: i) analytic or numerical population models or catalogs \citep[e.g.][]{Oguri2010,Collett2015}, ii) time delay measurement methods, especially linked to \h~measuring pipelines, where covariance between the delays and the mass model parameters can be provided, iii) machine learning methods that require light curves to study the accretion disc size \citep{Vernardos2019b}, microlensing events, and time delays, or static images to infer substructure \citep{Vernardos2020}.
We advocate that the MOLET software package can become an indispensable tool for data intensive future strong lensing science.

\section*{Data availability}
The data that support the findings of this study are openly available in github at \url{https://github.com/gvernard/molet}

\section*{Acknowledgements}
This project has received funding from the European Union's Horizon 2020 research and innovation programme under the Marie Sklodovska-Curie grant agreement No 897124.
The author has also received support from the Schmidt Futures foundation.
I would like to thank Chih-Fan Chen and Sherry Suyu for sharing their results and useful suggestions.
Thanks also to Eric Paic and Dominique Sluse for generating intrinsic light curve realizations of \rxj, and to Martin Millon and Justin Pierel for setting up the fits with \textsc{PyCS} and \textsc{SNTD} respectively.

\bibliographystyle{mnras}
\bibliography{biblio}

\begin{thebibliography}{}
\makeatletter
\relax
\def\mn@urlcharsother{\let\do\@makeother \do\$\do\&\do\#\do\^\do\_\do\%\do\~}
\def\mn@doi{\begingroup\mn@urlcharsother \@ifnextchar [ {\mn@doi@}
  {\mn@doi@[]}}
\def\mn@doi@[#1]#2{\def\@tempa{#1}\ifx\@tempa\@empty \href
  {http://dx.doi.org/#2} {doi:#2}\else \href {http://dx.doi.org/#2} {#1}\fi
  \endgroup}
\def\mn@eprint#1#2{\mn@eprint@#1:#2::\@nil}
\def\mn@eprint@arXiv#1{\href {http://arxiv.org/abs/#1} {{\tt arXiv:#1}}}
\def\mn@eprint@dblp#1{\href {http://dblp.uni-trier.de/rec/bibtex/#1.xml}
  {dblp:#1}}
\def\mn@eprint@#1:#2:#3:#4\@nil{\def\@tempa {#1}\def\@tempb {#2}\def\@tempc
  {#3}\ifx \@tempc \@empty \let \@tempc \@tempb \let \@tempb \@tempa \fi \ifx
  \@tempb \@empty \def\@tempb {arXiv}\fi \@ifundefined
  {mn@eprint@\@tempb}{\@tempb:\@tempc}{\expandafter \expandafter \csname
  mn@eprint@\@tempb\endcsname \expandafter{\@tempc}}}

\bibitem[\protect\citeauthoryear{Akiyama et~al.,}{Akiyama
  et~al.}{2019}]{EHT2019}
Akiyama K.,  et~al., 2019, \mn@doi [The Astrophysical Journal]
  {10.3847/2041-8213/ab1141}, 875, L6

\bibitem[\protect\citeauthoryear{Barkana}{Barkana}{1998}]{Barkana1998}
Barkana R.,  1998, \mn@doi [The Astrophysical Journal] {10.1086/305950}, 502,
  531

\bibitem[\protect\citeauthoryear{Bate \& Fluke}{Bate \& Fluke}{2012}]{Bate2012}
Bate N.~F.,  Fluke C.~J.,  2012, \mn@doi [The Astrophysical Journal]
  {10.1088/0004-637X/744/2/90}, 744, 90

\bibitem[\protect\citeauthoryear{Birrer \& Treu}{Birrer \&
  Treu}{2019}]{Birrer2019}
Birrer S.,  Treu T.,  2019, \mn@doi [Monthly Notices of the Royal Astronomical
  Society] {10.1093/mnras/stz2254}, 489, 2097

\bibitem[\protect\citeauthoryear{Bonvin, Chan, Millon, Rojas, Courbin, Chen,
  Fassnacht  \& Paic}{Bonvin et~al.}{2018}]{Bonvin2018}
Bonvin V.,  Chan J.~H.,  Millon M.,  Rojas K.,  Courbin F.,  Chen G.~C.,
  Fassnacht C.~D.,   Paic E.,  2018, \mn@doi [Astronomy and Astrophysics]
  {10.1051/0004-6361/201833287}, 616, 1

\bibitem[\protect\citeauthoryear{Cackett, Horne  \& Winkler}{Cackett
  et~al.}{2007}]{Cackett2007}
Cackett E.~M.,  Horne K.,   Winkler H.,  2007, \mn@doi [Monthly Notices of the
  Royal Astronomical Society] {10.1111/j.1365-2966.2007.12098.x}, 380, 669

\bibitem[\protect\citeauthoryear{Chan, Schive, Wong  \& Chiueh}{Chan
  et~al.}{2020}]{Chan2020}
Chan J. H.~H.,  Schive H.-Y.,  Wong S.-K.,   Chiueh T.,  2020, preprint
  (astro-ph/2002.10473), pp~1--6

\bibitem[\protect\citeauthoryear{Chen et~al.,}{Chen et~al.}{2016}]{Chen2016}
Chen G. C.~F.,  et~al., 2016, \mn@doi [Monthly Notices of the Royal
  Astronomical Society] {10.1093/mnras/stw991}, 462, 3457

\bibitem[\protect\citeauthoryear{Chen et~al.,}{Chen et~al.}{2018}]{Chen2018}
Chen G.~C.,  et~al., 2018, \mn@doi [Monthly Notices of the Royal Astronomical
  Society] {10.1093/MNRAS/STY2350}, 481, 1115

\bibitem[\protect\citeauthoryear{Collett}{Collett}{2015}]{Collett2015}
Collett T.,  2015, \mn@doi [The Astrophysical Journal]
  {10.1088/0004-637X/811/1/20}, 811, 20

\bibitem[\protect\citeauthoryear{Collett \& Smith}{Collett \&
  Smith}{2020}]{Collett2020}
Collett T.~E.,  Smith R.~J.,  2020, Monthly Notices of the Royal Astronomical
  Society, 497, 1654

\bibitem[\protect\citeauthoryear{Cornachione, Morgan, Millon, Bentz, Courbin,
  Bonvin  \& Falco}{Cornachione et~al.}{2020}]{Cornachione2020}
Cornachione M.~A.,  Morgan C.~W.,  Millon M.,  Bentz M.~C.,  Courbin F.,
  Bonvin V.,   Falco E.~E.,  2020, \mn@doi [The Astrophysical Journal]
  {10.3847/1538-4357/ab557a}, 895, 125

\bibitem[\protect\citeauthoryear{Dalal \& Kochanek}{Dalal \&
  Kochanek}{2002}]{Dalal2002}
Dalal N.,  Kochanek C.~S.,  2002, The Astrophysical Journal, 572, 25

\bibitem[\protect\citeauthoryear{Despali, Vegetti, White, Giocoli  \& van~den
  Bosch}{Despali et~al.}{2018}]{Despali2018}
Despali G.,  Vegetti S.,  White S.~D.,  Giocoli C.,   van~den Bosch F.~C.,
  2018, \mn@doi [Monthly Notices of the Royal Astronomical Society]
  {10.1093/mnras/sty159}, 475, 5424

\bibitem[\protect\citeauthoryear{Fausnaugh, Peterson, Starkey, Horne  \& {AGN
  Storm Collaboration}}{Fausnaugh et~al.}{2017}]{Fausnaugh2017}
Fausnaugh M.~M.,  Peterson B.~M.,  Starkey D.~A.,  Horne K.,   {AGN Storm
  Collaboration} 2017, \mn@doi [Frontiers in Astronomy and Space Sciences]
  {10.3389/fspas.2017.00055}, 4, 55

\bibitem[\protect\citeauthoryear{Fluke \& Webster}{Fluke \&
  Webster}{1999}]{Fluke1999}
Fluke C.~J.,  Webster R.~L.,  1999, \mn@doi [Monthly Notices of the Royal
  Astronomical Society] {10.1046/j.1365-8711.1999.02109.x}, 302, 68

\bibitem[\protect\citeauthoryear{Foxley-Marrable, Collett, Vernardos, Goldstein
   \& Bacon}{Foxley-Marrable et~al.}{2018}]{FoxleyMarrable2018}
Foxley-Marrable M.,  Collett T.~E.,  Vernardos G.,  Goldstein D.~A.,   Bacon
  D.,  2018, Monthly Notices of the Royal Astronomical Society, 478, 5081

\bibitem[\protect\citeauthoryear{Freedman}{Freedman}{2017}]{Freedman2017}
Freedman W.~L.,  2017, \mn@doi [Nature Astronomy] {10.1038/s41550-017-0121}, 1,
  121

\bibitem[\protect\citeauthoryear{Gavazzi, Treu, Koopmans, Bolton, Moustakas,
  Burles  \& Marshall}{Gavazzi et~al.}{2008}]{Gavazzi2008}
Gavazzi R.,  Treu T.,  Koopmans L. V.~E.,  Bolton A.~S.,  Moustakas L.~A.,
  Burles S.,   Marshall P.~J.,  2008, \mn@doi [The Astrophysical Journal]
  {10.1086/529541}, 677, 1046

\bibitem[\protect\citeauthoryear{Gilman, Birrer, Nierenberg, Treu, Du  \&
  Benson}{Gilman et~al.}{2020a}]{Gilman2020a}
Gilman D.,  Birrer S.,  Nierenberg A.,  Treu T.,  Du X.,   Benson A.,  2020a,
  \mn@doi [Monthly Notices of the Royal Astronomical Society]
  {10.1093/mnras/stz3480}, 491, 6077

\bibitem[\protect\citeauthoryear{Gilman, Birrer  \& Treu}{Gilman
  et~al.}{2020b}]{Gilman2020b}
Gilman D.,  Birrer S.,   Treu T.,  2020b, Astronomy \& Astrophysics, 642, A194

\bibitem[\protect\citeauthoryear{Goldstein, Nugent, Kasen  \&
  Collett}{Goldstein et~al.}{2018}]{Goldstein2018b}
Goldstein D.~A.,  Nugent P.~E.,  Kasen D.~N.,   Collett T.~E.,  2018, \mn@doi
  [The Astrophysical Journal] {10.3847/1538-4357/aaa975}, 855, 22

\bibitem[\protect\citeauthoryear{Guy et~al.,}{Guy et~al.}{2007}]{Guy2007}
Guy J.,  et~al., 2007, \mn@doi [Astronomy and Astrophysics]
  {10.1051/0004-6361:20066930}, 466, 11

\bibitem[\protect\citeauthoryear{Harvey, Valkenburg, Tamone, Boyarsky, Courbin
  \& Lovell}{Harvey et~al.}{2020}]{Harvey2020}
Harvey D.,  Valkenburg W.,  Tamone A.,  Boyarsky A.,  Courbin F.,   Lovell M.,
  2020, \mn@doi [Monthly Notices of the Royal Astronomical Society]
  {10.1093/mnras/stz3305}, 491, 4247

\bibitem[\protect\citeauthoryear{Huber et~al.,}{Huber et~al.}{2019}]{Huber2019}
Huber S.,  et~al., 2019, Astronomy \& Astrophysics, 631, A161

\bibitem[\protect\citeauthoryear{Kayser, Refsdal  \& Stabell}{Kayser
  et~al.}{1986}]{Kayser1986}
Kayser R.,  Refsdal S.,   Stabell R.,  1986, Astronomy \& Astrophysics, 166, 36

\bibitem[\protect\citeauthoryear{Keeton}{Keeton}{2001}]{Keeton2001b}
Keeton C.~R.,  2001, preprint (astro-ph/0102341)

\bibitem[\protect\citeauthoryear{Keeton}{Keeton}{2010}]{Keeton2010}
Keeton C.~R.,  2010, \mn@doi [General Relativity and Gravitation]
  {10.1007/s10714-010-1041-1}, 42, 2151

\bibitem[\protect\citeauthoryear{Keeton \& Moustakas}{Keeton \&
  Moustakas}{2009}]{Keeton2009}
Keeton C.~R.,  Moustakas L.~A.,  2009, \mn@doi [Astrophysical Journal]
  {10.1088/0004-637X/699/2/1720}, 699, 1720

\bibitem[\protect\citeauthoryear{Kessler et~al.,}{Kessler
  et~al.}{2009}]{Kessler2009}
Kessler R.,  et~al., 2009, \mn@doi [Publications of the Astronomical Society of
  the Pacific] {10.1086/605984}, 121, 1028

\bibitem[\protect\citeauthoryear{Kochanek}{Kochanek}{2004}]{Kochanek2004}
Kochanek C.~S.,  2004, \mn@doi [The Astrophysical Journal] {10.1086/382180},
  605, 58

\bibitem[\protect\citeauthoryear{Komatsu et~al.,}{Komatsu
  et~al.}{2011}]{Komatsu2011}
Komatsu E.,  et~al., 2011, \mn@doi [Astrophysical Journal, Supplement Series]
  {10.1088/0067-0049/192/2/18}, 192, 18

\bibitem[\protect\citeauthoryear{Kormann, Schneider  \& Bartelmann}{Kormann
  et~al.}{1994}]{Kormann1994}
Kormann R.,  Schneider P.,   Bartelmann M.,  1994, Astronomy \& Astrophysics,
  284, 285

\bibitem[\protect\citeauthoryear{Kundic \& Wambsganss}{Kundic \&
  Wambsganss}{1993}]{Kundic1993}
Kundic T.,  Wambsganss J.,  1993, \mn@doi [The Astrophysical Journal]
  {10.1192/bjp.111.479.1009-a}, 404, 455

\bibitem[\protect\citeauthoryear{{LSST Science Collaborations}}{{LSST Science
  Collaborations}}{2009}]{LSST2009}
{LSST Science Collaborations} 2009, preprint (astro-ph/0912.0201)

\bibitem[\protect\citeauthoryear{Lage}{Lage}{2019}]{Lage2019}
Lage C.,  2019, preprint (astro-ph/1911.09577)

\bibitem[\protect\citeauthoryear{Lewis \& Irwin}{Lewis \&
  Irwin}{1995}]{Lewis1995}
Lewis G.~F.,  Irwin M.~J.,  1995, Monthly Notices of the Royal Astronomical
  Society, 276, 103

\bibitem[\protect\citeauthoryear{Liao}{Liao}{2019}]{Liao2019}
Liao K.,  2019, \mn@doi [The Astrophysical Journal] {10.3847/1538-4357/aaf733},
  871, 113

\bibitem[\protect\citeauthoryear{MacLeod et~al.,}{MacLeod
  et~al.}{2010}]{MacLeod2010}
MacLeod C.~L.,  et~al., 2010, \mn@doi [The Astrophysical Journal]
  {10.1088/0004-637X/721/2/1014}, 721, 1014

\bibitem[\protect\citeauthoryear{MacLeod, Jones, Agol  \& Kochanek}{MacLeod
  et~al.}{2013}]{MacLeod2013}
MacLeod C.~L.,  Jones R.,  Agol E.,   Kochanek C.~S.,  2013, \mn@doi [The
  Astrophysical Journal] {10.1088/0004-637X/773/1/35}, 773, 35

\bibitem[\protect\citeauthoryear{Metcalf \& Zhao}{Metcalf \&
  Zhao}{2002}]{Metcalf2002}
Metcalf R.~B.,  Zhao H.,  2002, The Astrophysical Journal Letters, 567, L5

\bibitem[\protect\citeauthoryear{Millon, Tewes, Bonvin, Lengen  \&
  Courbin}{Millon et~al.}{2020a}]{Millon2020c}
Millon M.,  Tewes M.,  Bonvin V.,  Lengen B.,   Courbin F.,  2020a, \mn@doi
  [Journal of Open Source Software] {10.21105/joss.02654}, 5, 2654

\bibitem[\protect\citeauthoryear{Millon et~al.,}{Millon
  et~al.}{2020b}]{Millon2020b}
Millon M.,  et~al., 2020b, \mn@doi [Astronomy and Astrophysics]
  {10.1051/0004-6361/201937351}, 639, 1

\bibitem[\protect\citeauthoryear{Millon et~al.,}{Millon
  et~al.}{2020c}]{Millon2020a}
Millon M.,  et~al., 2020c, \mn@doi [Astronomy \& Astrophysics]
  {10.1051/0004-6361/202037740}, 640, A105

\bibitem[\protect\citeauthoryear{Morgan et~al.,}{Morgan
  et~al.}{2012}]{Morgan2012}
Morgan C.~W.,  et~al., 2012, \mn@doi [The Astrophysical Journal]
  {10.1088/0004-637X/756/1/52}, 756, 52

\bibitem[\protect\citeauthoryear{Morgan, Hyer, Bonvin, Mosquera, Cornachione,
  Courbin, Kochanek  \& Falco}{Morgan et~al.}{2018}]{Morgan2018}
Morgan C.~W.,  Hyer G.~E.,  Bonvin V.,  Mosquera A.~M.,  Cornachione M.,
  Courbin F.,  Kochanek C.~S.,   Falco E.~E.,  2018, \mn@doi [The Astrophysical
  Journal] {10.3847/1538-4357/aaed3e}, 869, 106

\bibitem[\protect\citeauthoryear{Mortonson, Schechter  \& Wambsganss}{Mortonson
  et~al.}{2005}]{Mortonson2005}
Mortonson M.~J.,  Schechter P.~L.,   Wambsganss J.,  2005, \mn@doi [The
  Astrophysical Journal] {10.1086/431195}, 628, 594

\bibitem[\protect\citeauthoryear{Mosquera \& Kochanek}{Mosquera \&
  Kochanek}{2011}]{Mosquera2011b}
Mosquera A.~M.,  Kochanek C.~S.,  2011, \mn@doi [The Astrophysical Journal]
  {10.1088/0004-637X/738/1/96}, 738, 96

\bibitem[\protect\citeauthoryear{Mukherjee et~al.,}{Mukherjee
  et~al.}{2018}]{Mukherjee2018}
Mukherjee S.,  et~al., 2018, \mn@doi [Monthly Notices of the Royal Astronomical
  Society] {10.1093/mnras/sty1741}, 479, 4108

\bibitem[\protect\citeauthoryear{Neira, Anguita  \& Vernardos}{Neira
  et~al.}{2020}]{Neira2020}
Neira F.,  Anguita T.,   Vernardos G.,  2020, \mn@doi [Monthly Notices of the
  Royal Astronomical Society] {10.1093/mnras/staa1208}, 495, 544

\bibitem[\protect\citeauthoryear{Nierenberg et~al.,}{Nierenberg
  et~al.}{2017}]{Nierenberg2017}
Nierenberg A.~M.,  et~al., 2017, \mn@doi [Monthly Notices of the Royal
  Astronomical Society] {10.1093/mnras/stx1400}, 471, 2224

\bibitem[\protect\citeauthoryear{Oguri}{Oguri}{2019}]{Oguri2019}
Oguri M.,  2019, \mn@doi [Reports on Progress in Physics]
  {10.1088/1361-6633/ab4fc5}, 82

\bibitem[\protect\citeauthoryear{Oguri \& Marshall}{Oguri \&
  Marshall}{2010}]{Oguri2010}
Oguri M.,  Marshall P.~J.,  2010, \mn@doi [Monthly Notices of the Royal
  Astronomical Society] {10.1111/j.1365-2966.2010.16639.x}, 405, 2579

\bibitem[\protect\citeauthoryear{Padovani et~al.,}{Padovani
  et~al.}{2017}]{Padovani2017}
Padovani P.,  et~al., 2017, \mn@doi [Astronomy and Astrophysics Review]
  {10.1007/s00159-017-0102-9}, 25

\bibitem[\protect\citeauthoryear{Paic}{Paic}{2021}]{Paic2021}
Paic E.,  2021, in preparation

\bibitem[\protect\citeauthoryear{Petkova, Metcalf  \& Giocoli}{Petkova
  et~al.}{2014}]{Petkova2014}
Petkova M.,  Metcalf R.~B.,   Giocoli C.,  2014, \mn@doi [Monthly Notices of
  the Royal Astronomical Society] {10.1093/mnras/stu1860}, 445, 1954

\bibitem[\protect\citeauthoryear{{Planck Collaboration} et~al.,}{{Planck
  Collaboration} et~al.}{2020}]{PlanckCollaboration2018vi}
{Planck Collaboration} P.,  et~al., 2020, Astronomy \& Astrophysics, 641, A6

\bibitem[\protect\citeauthoryear{Plazas}{Plazas}{2020}]{Plazas2020}
Plazas A.~A.,  2020, \mn@doi [Symmetry] {10.3390/SYM12040494}, 12

\bibitem[\protect\citeauthoryear{Poindexter \& Kochanek}{Poindexter \&
  Kochanek}{2010}]{Poindexter2010a}
Poindexter S.,  Kochanek C.~S.,  2010, \mn@doi [The Astrophysical Journal]
  {10.1088/0004-637X/712/1/658}, 712, 658

\bibitem[\protect\citeauthoryear{Riess et~al.,}{Riess et~al.}{2016}]{Riess2016}
Riess A.~G.,  et~al., 2016, \mn@doi [The Astrophysical Journal]
  {10.3847/0004-637x/826/1/56}, 826, 56

\bibitem[\protect\citeauthoryear{Rusu et~al.,}{Rusu et~al.}{2017}]{Rusu2017}
Rusu C.~E.,  et~al., 2017, \mn@doi [Monthly Notices of the Royal Astronomical
  Society] {10.1093/mnras/stx285}, 467, 4220

\bibitem[\protect\citeauthoryear{Sartori, Trakhtenbrot, Schawinski, Caplar,
  Treister  \& Zhang}{Sartori et~al.}{2019}]{Sartori2019}
Sartori L.~F.,  Trakhtenbrot B.,  Schawinski K.,  Caplar N.,  Treister E.,
  Zhang C.,  2019, \mn@doi [The Astrophysical Journal]
  {10.3847/1538-4357/ab3c55}, 883, 139

\bibitem[\protect\citeauthoryear{Schechter \& Wambsganss}{Schechter \&
  Wambsganss}{2002}]{Schechter2002}
Schechter P.~L.,  Wambsganss J.,  2002, \mn@doi [The Astrophysical Journal]
  {10.1086/343856}, 580, 685

\bibitem[\protect\citeauthoryear{Schmidt \& Wambsganss}{Schmidt \&
  Wambsganss}{2010}]{Schmidt2010}
Schmidt R.~W.,  Wambsganss J.,  2010, \mn@doi [Gen. Relativ. and Gravit.]
  {10.1007/s10714-010-0956-x}, 42, 2127

\bibitem[\protect\citeauthoryear{Schneider, Kochanek  \& Wambsganss}{Schneider
  et~al.}{2006}]{Schneider2006}
Schneider P.,  Kochanek C.~S.,   Wambsganss J.,  2006, in Meylan G.,  Jetzer
  P.,   North P.,  eds, Saas-Fee Advanced Course vol. 33. Springer, Berlin

\bibitem[\protect\citeauthoryear{Sersic}{Sersic}{1963}]{Sersic1963}
Sersic J.~L.,  1963, Bolet{\'{i}}n de la Asociaci{\'{o}}n Argentina de
  Astronom{\'{i}}a, 6, 41

\bibitem[\protect\citeauthoryear{Shajib et~al.,}{Shajib
  et~al.}{2019}]{Shajib2019}
Shajib A.~J.,  et~al., 2019, \mn@doi [Monthly Notices of the Royal Astronomical
  Society] {10.1093/mnras/sty3397}, 483, 5649

\bibitem[\protect\citeauthoryear{Shakura \& Sunyaev}{Shakura \&
  Sunyaev}{1973}]{Shakura1973}
Shakura N.~I.,  Sunyaev R.~A.,  1973, Astronomy \& Astrophysics, 24, 337

\bibitem[\protect\citeauthoryear{Sluse \& Tewes}{Sluse \&
  Tewes}{2014}]{Sluse2014}
Sluse D.,  Tewes M.,  2014, \mn@doi [Astronomy and Astrophysics]
  {10.1051/0004-6361/201424776}, 571, 1

\bibitem[\protect\citeauthoryear{Sluse, Claeskens, Hutsem{\'{e}}kers  \&
  Surdej}{Sluse et~al.}{2006}]{Sluse2006}
Sluse D.,  Claeskens J.-F.,  Hutsem{\'{e}}kers D.,   Surdej J.,  2006, \mn@doi
  [Astronomy \& Astrophysics] {10.1051/0004-6361:20066821}, 449, 539

\bibitem[\protect\citeauthoryear{Stacey \& McKean}{Stacey \&
  McKean}{2018}]{Stacey2018}
Stacey H.~R.,  McKean J.~P.,  2018, \mn@doi [Monthly Notices of the Royal
  Astronomical Society: Letters] {10.1093/mnrasl/sly153}, 481, L40

\bibitem[\protect\citeauthoryear{Suyu et~al.,}{Suyu et~al.}{2013}]{Suyu2013}
Suyu S.~H.,  et~al., 2013, \mn@doi [The Astrophysical Journal]
  {10.1088/0004-637X/766/2/70}, 766, 70

\bibitem[\protect\citeauthoryear{Suyu, Huber, Ca{\~{n}}ameras, Kromer, Schuldt,
  Taubenberger, Yıldırım  \& Bonvin}{Suyu et~al.}{2020}]{Suyu2020}
Suyu S.~H.,  Huber S.,  Ca{\~{n}}ameras R.,  Kromer M.,  Schuldt S.,
  Taubenberger S.,  Yıldırım A.,   Bonvin V.,  2020, Astronomy \&
  Astrophysics, 644, A162

\bibitem[\protect\citeauthoryear{Tewes, Courbin  \& Meylan}{Tewes
  et~al.}{2013}]{Tewes2013}
Tewes M.,  Courbin F.,   Meylan G.,  2013, \mn@doi [Astronomy \& Astrophysics]
  {10.1051/0004-6361/201220123}, 553, 120

\bibitem[\protect\citeauthoryear{Tie \& Kochanek}{Tie \&
  Kochanek}{2018}]{Tie2018a}
Tie S.~S.,  Kochanek C.~S.,  2018, \mn@doi [Monthly Notices of the Royal
  Astronomical Society] {10.1093/mnras/stx2348}, 473, 80

\bibitem[\protect\citeauthoryear{Treu \& Marshall}{Treu \&
  Marshall}{2016}]{Treu2016}
Treu T.,  Marshall P.~J.,  2016, \mn@doi [Astronomy and Astrophysics Review]
  {10.1007/s00159-016-0096-8}, 24

\bibitem[\protect\citeauthoryear{Vernardos}{Vernardos}{2019}]{Vernardos2019a}
Vernardos G.,  2019, Monthly Notices of the Royal Astronomical Society, 483,
  5583

\bibitem[\protect\citeauthoryear{Vernardos \& Fluke}{Vernardos \&
  Fluke}{2013}]{Vernardos2013}
Vernardos G.,  Fluke C.~J.,  2013, Monthly Notices of the Royal Astronomical
  Society, 434, 832

\bibitem[\protect\citeauthoryear{Vernardos \& Fluke}{Vernardos \&
  Fluke}{2014}]{Vernardos2014b}
Vernardos G.,  Fluke C.~J.,  2014, Astronomy \& Computing, 6, 1

\bibitem[\protect\citeauthoryear{Vernardos \& Tsagkatakis}{Vernardos \&
  Tsagkatakis}{2019}]{Vernardos2019b}
Vernardos G.,  Tsagkatakis G.,  2019, \mn@doi [Monthly Notices of the Royal
  Astronomical Society] {10.1093/mnras/stz868}, 486, 1944

\bibitem[\protect\citeauthoryear{Vernardos, Fluke, Bate  \& Croton}{Vernardos
  et~al.}{2014}]{Vernardos2014a}
Vernardos G.,  Fluke C.~J.,  Bate N.~F.,   Croton D.~J.,  2014, The
  Astrophysical Journal Supplement Series, 211, 16

\bibitem[\protect\citeauthoryear{Vernardos, Fluke, Bate, Croton  \&
  Vohl}{Vernardos et~al.}{2015}]{Vernardos2015}
Vernardos G.,  Fluke C.~J.,  Bate N.~F.,  Croton D.~J.,   Vohl D.,  2015, The
  Astrophysical Journal Supplement Series, 217, 23

\bibitem[\protect\citeauthoryear{Vernardos, Tsagkatakis  \& Pantazis}{Vernardos
  et~al.}{2020}]{Vernardos2020}
Vernardos G.,  Tsagkatakis G.,   Pantazis Y.,  2020, \mn@doi [Monthly Notices
  of the Royal Astronomical Society] {10.1093/mnras/staa3201}, 499, 5641

\bibitem[\protect\citeauthoryear{Wambsganss, Paczynski  \&
  Schneider}{Wambsganss et~al.}{1990}]{Wambsganss1990b}
Wambsganss J.,  Paczynski B.,   Schneider P.,  1990, The Astrophysical Journal,
  358, L33

\bibitem[\protect\citeauthoryear{Wong et~al.,}{Wong et~al.}{2017}]{Wong2017}
Wong K.~C.,  et~al., 2017, \mn@doi [Monthly Notices of the Royal Astronomical
  Society] {10.1093/mnras/stw3077}, 465, 4895

\bibitem[\protect\citeauthoryear{Wong, Suyu, Chen, Rusu, Millon, Sluse, Bonvin
  \& Fassnacht}{Wong et~al.}{2020}]{Wong2020}
Wong K.~C.,  Suyu S.~H.,  Chen G. C.-F.,  Rusu C.~E.,  Millon M.,  Sluse D.,
  Bonvin V.,   Fassnacht C.~D.,  2020, \mn@doi [Monthly Notices of the Royal
  Astronomical Society] {10.1093/mnras/stz3094}, 498, 1420

\bibitem[\protect\citeauthoryear{Wyithe \& Turner}{Wyithe \&
  Turner}{2001}]{Wyithe2001}
Wyithe J. S.~B.,  Turner E.~L.,  2001, \mn@doi [Monthly Notices of the Royal
  Astronomical Society] {10.1046/j.1365-8711.2001.03917.x}, 320, 21

\bibitem[\protect\citeauthoryear{Wyithe, Webster  \& Turner}{Wyithe
  et~al.}{2000}]{Wyithe2000a}
Wyithe J. S.~B.,  Webster R.~L.,   Turner E.~L.,  2000, Monthly Notices of the
  Royal Astronomical Society, 312, 843

\bibitem[\protect\citeauthoryear{de Grijs, Courbin,
  Mart{\'{i}}nez-V{\'{a}}zquez, Monelli, Oguri  \& Suyu}{de~Grijs
  et~al.}{2017}]{deGrijs2017}
de Grijs R.,  Courbin F.,  Mart{\'{i}}nez-V{\'{a}}zquez C.~E.,  Monelli M.,
  Oguri M.,   Suyu S.~H.,  2017, \mn@doi [Space Science Reviews]
  {10.1007/s11214-017-0395-z}, 212, 1743

\makeatother
\end{thebibliography}

\end{document}